\documentclass[aps,prb,twocolumn,showpacs,superscriptaddress]{revtex4}

\usepackage{graphicx}

\newcommand{\bvec}[1]{{\bf #1}}

 \newcommand{\bv}[1]{{\bf #1}}

\newcommand{\bra}[1]{\left< #1 \right|} 

\newcommand{\ket}[1]{\left|#1 \right>}

\newcommand{\EqLabel}[1] { \label{#1} }

\newlength\FigWidth \FigWidth 80 true mm

\begin{document}

\title{Effects of large disorder on the Hofstadter butterfly}

\author{Chenggang Zhou} \affiliation{Department of Electrical
Engineering, Princeton University, Princeton NJ 08544, USA}

\author{Mona Berciu} \affiliation{Department of Physics and Astronomy,
University of British Columbia, Vancouver, BC V6T 1Z1, Canada }

\author{R. N. Bhatt} \affiliation{Department of Electrical
Engineering, Princeton University, Princeton NJ 08544, USA}

\date{\today}

\begin{abstract}
Motivated by the recent experiments on periodically modulated, two
dimensional electron systems 
placed in large transversal 
magnetic fields, we investigate the interplay between the effects of
disorder and periodic potentials in the integer quantum Hall
regime. In particular, we study the case where disorder is larger than
the periodic modulation, but both are small enough that Landau level
mixing is negligible. In this limit, the self-consistent Born
approximation is inadequate. We carry extensive numerical calculations
to understand the relevant physics in the lowest Landau level, such as
the spectrum and nature (localized or extended) of the
wave functions. Based on our results, we propose a
qualitative explanation of the new features uncovered recently in
transport measurements.
\end{abstract}

\maketitle

\section{Introduction} 
\label{sec1}

Two-dimensional electron systems (2DES) placed in a uniform
perpendicular magnetic field exhibit a rich variety of phenomena, such
as the integer\cite{IQHE} and fractional\cite{FQHE} quantum Hall
effects. \cite{Springer} Another well-studied problem is that of a
2DES in a uniform perpendicular magnetic field subjected to a periodic
potential. Even before the discovery of the quantum Hall effects,
Hofstadter\cite{Hofst} showed that in this case, the electronic bands
split into a remarkable fractal structure of subbands and gaps, the
so-called Hofstadter butterfly. Two ``asymptotic'' regimes are usually
considered: (i) if the magnitude of the periodic potential is very
large compared to the cyclotron energy and the Zeeman splitting, then
one can use lattice models to describe the hopping of electrons
between Wannier-like states localized at the minima of the periodic
potential, whereas (ii) if the magnitude of the periodic potential is
small compared to the cyclotron energy, the periodic potential lifts
the degeneracy of each Landau level. In both cases, the resulting butterfly
structure is a function only of the ratio between the flux
$\phi=B{\cal A}$ of the magnetic field through the unit cell of the
periodic lattice, and the elementary magnetic flux
$\phi_0=hc/e$. Remarkably, if $\phi/\phi_0$ of the first asymptotic
case is equal to $\phi_0/\phi$ of the second case, their
electronic structures are solutions of the same Harper's
equation.\cite{equiv} If the periodic potential is comparable to the
cyclotron energy, Landau level mixing must be taken into account;
although Landau levels still split into subbands, the structure is no
longer universal, but depends also on the ratio of the periodic
potential amplitude and the cyclotron energy.\cite{germans}

Experimentally, the case with a small periodic modulation can be
realized more easily. This is because the periodic potential is
usually imprinted at some distance from the 2DES layer; as a result,
its magnitude in the 2DES is considerably attenuated. The interesting
cases to study experimentally also correspond to small values of
$\phi/\phi_0$ (of order unity), where the butterfly structure shows a
small number of subbands separated by large gaps, and should therefore
be easier to identify. Periodic modulations have been created using
lithography\cite{Holland,Weimann,Weiss} and holographic
illumination. \cite{Wulf} The lattice constants of the resulting
square lattices are of order 100~nm. As a result,
the condition $\phi/\phi_0 \approx 2$ (for instance) is satisfied for
$B\approx 0.8$~T. This is a very low value, in the Shubnikov-de Haas
(SdH) regime, not the high-$B$ quantum regime. Significant Landau
level mixing and complications from the fact that the Fermi level is
inside one of the higher Landau levels for such small $B$-values make
the identification of the Hofstadter structure difficult.

Recently, a new method for lateral periodic modulation has been
developed using a self-organized ordered phase of a diblock copolymer
deposited on a GaAs/AlGaAs heterostructure.\cite{Sorin} The polymer
spheres create a 2D triangular lattice with a lattice constant of
about 39~nm. The corresponding unit cell area is almost an order of
magnitude smaller than those achieved in previous experiments,
implying that the condition $\phi/\phi_0 \approx 2$ is now satisfied
for very strong magnetic fields, $B\approx 6$~T. At such high magnetic
fields the system is in the strong quantum regime, and Landau level
mixing can be safely ignored. For the experimental 2DES electron
concentrations, the Fermi level is in the spin-down lowest Landau
level.\cite{Sorin} As a result, this experimental setup appears more
promising for the successful observation of the butterfly.

Nevertheless, one must take into account the disorder which is present
in the system (without disorder, there is no integer Quantum Hall
Effect -- IQHE -- to begin with). If
the disorder is very small compared to the periodic potential
amplitude, one expects that the subbands of the Hofstadter structure
are ``smeared'' on a scale $\hbar/\tau$, where $\tau$ is the
scattering time, and $\tau \rightarrow \infty$ as disorder becomes
vanishingly small. As a result, the larger gaps in the Hofstadter
structure should remain open at the positions predicted in the absence
of disorder, and one expects a series of minima in the longitudinal
conductivity as the Fermi level traverses such gaps. The experiment
indeed shows a very non-trivial modification of the longitudinal
resistivity, with many peaks and valleys appearing in what is (in the
absence of the periodic modulation) a smooth Lorentz-like
peak.\cite{Sorin} However, the position of the minima in $\rho_{xx}$
do not track the positions of the main gaps in the corresponding
Hofstadter butterfly structure. Instead, the data suggests that in
this experimental setup, disorder is not small, but rather large
compared to the estimated amplitude of the periodic potential. This is
not a consequence of poor samples, since these 2DES have high
mobilities. It is due to the fact that the periodic modulation is
considerably attenuated in the 2DES, leading to a small energy scale
for the Hofstadter butterfly spectrum as compared to $\hbar/\tau$. As
a result, the Hofstadter structure predicted in the absence of
disorder is of little use for interpreting the experimental data. One
might expect that in this case the periodic potential should have
basically no effect on the disorder-broadened Landau level. This is
indeed true for the strongly localized states at the top and bottom of
the Landau level. However, states in the center of the Landau level
extend over many unit cells of the periodic potential, and, as we
demonstrate in the following, are non-trivially modified by its
presence.

In this paper, we investigate numerically the behavior of a 2DES
subject to a perpendicular magnetic field, a periodic potential and a
disorder potential, under conditions applicable to the experimental
system. The effective electron mass in GaAs is $0.067m_e$ while the
magnetic fields of interest are on the order of 10~T. Under these
conditions, the cyclotron energy $\hbar \omega_c$, of the order of
200~K, is the largest energy scale in the problem.
The Zeeman energy $g^*\mu_B B$ for these fields is roughly 3~K, but
electron interaction effects lead to a considerable enhancement of
the spin splitting between the (spin polarized) Landau levels, which
has been measured to be 20~K.\cite{Sorin2} The amplitude of the
periodic potential's largest Fourier components is estimated to be of
the order of 1~K, and the scattering rate from the known zero field
mobility is estimated to be $\hbar/\tau \sim 8$~K. \cite{Chaikin} As a
result of this ordering of energy scales, we neglect Landau level
inter-mixing and study non-perturbatively the combined effects of a
periodic and a large smooth disorder potential on the electronic
structure of the lowest Landau level. Previously, the effects of
small disorder on a Hofstadter butterfly have been perturbatively
investigated using the self-consistent Born approximation (SCBA),
\cite{MacDonald} and the combined effect of white-noise disorder and
periodic modulation on Hall resistance was studied following the
scaling theory of IQHE.\cite{Huckestein} Our results reveal
details of the electronic structure not investigated previously.

\begin{figure}[t]
\includegraphics[width=0.9\FigWidth]{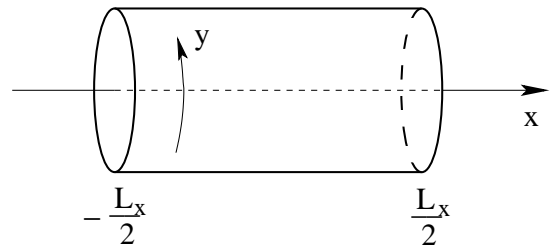}
\caption{ The two-lead geometry considered: the
 finite-size 2DES has periodic boundary conditions in the
 $y$-direction, and is attached to metallic leads at the $x=\pm L_x/2$
 ends. }
\label{fig1}
\end{figure}

The two-lead geometry we consider is schematically shown in
Fig.~\ref{fig1}: the finite 2DES is assumed to have periodic boundary
conditions in the $y$-direction (along which the Hall currents flow),
and is connected to metallic leads at the $x=-L_x/2$ and $x=+L_x/2$
edges. In particular, in this paper we study the effects of the
periodic potential on the extended states carrying longitudinal
currents between the two leads, and identify a number of interesting
properties, in qualitative agreement with simple arguments provided by
a semi-classical picture. Our main conclusion is that while the
beautiful Hofstadter structure is destroyed by large disorder, the
system still exhibits very interesting and non-trivial physics.

The paper is organized as follows: in Section~\ref{sec2} we briefly
review the computation of the Hofstadter structure for a
small-amplitude periodic potential. In Section~\ref{sec3} we describe
the type of disorder potentials considered. Section~\ref{sec4}
describes the numerical methods used to analyze the spectrum and the
nature of the electronic states, with both semi-classical and fully
quantum-mechanical formalisms. Results are presented in Section~\ref{sec5}, 
while Section~\ref{sec6} contains discussions and
 a summary of our conclusions.

\section{Periodic Potential}
\label{sec2}

To clarify our notation, we briefly review the problem of a free
electron of charge $-e$ moving in a 2D plane (from now on, the
$xy$-plane, of dimension $L_x \times L_y$) in a magnetic field
$\bv{B}=B\bv{e}_z$ perpendicular to the plane, as described by
$$ {\cal H} = { 1 \over 2m} \left(\bv{p}+{e \over c}\bv{A}\right)^2 -
{1 \over 2} g\mu_B\vec{\sigma}\cdot \bv{B}
$$ In the Landau gauge $\bvec{A} = (0, Bx,0)$, the eigenfunctions of
the Schr\"odinger equation ${\cal H}|n,k_y,\sigma\rangle=
E_{n,\sigma}|n,k_y,\sigma\rangle$ are: 
\begin{equation}
\EqLabel{2.1} \langle \bvec r | n, k_y, \sigma \rangle = { e^{-ik_yy}
\over \sqrt{L_y}} e^{-{1 \over 2}\left( {x \over l} - lk_y\right)^2 }
{ H_n\left({x \over l} - lk_y\right) \over
\sqrt{2^nn!\sqrt{\pi}l}}\chi_\sigma,
\end{equation}
with eigenenergies
\begin{equation}
\EqLabel{2.2} E_{n,\sigma} = \hbar\omega_c\left( n + { 1 \over2
}\right) - {1 \over2} g \mu_B B\sigma.
\end{equation}
Here $l= \sqrt{ \hbar c/ eB}$ is the magnetic length, $\omega_c ={eB/
mc}$ is the cyclotron frequency, $H_n(x)$ are the Hermite 
polynomials and $\chi_{+1}^T =(1\; 0)$, respectively $\chi_{-1}^T
=(0\; 1)$ are the eigenspinors of $\sigma_z$: $\sigma_z \chi_\sigma
=\sigma \chi_\sigma$. 

The degeneracy of a Landau level is given by the number of distinct
$k_y$ values allowed. Imposing cyclic boundary conditions in the
$y$-direction, we find
\begin{equation}
\EqLabel{2.3} k_y = { 2\pi \over L_y} j,
\end{equation}
where $j$ is an integer. The allowed values for $j$ are found from the
condition that the electron wave-functions, which are centered at
positions $x_j = l^2 k_y = l^2 2\pi j/L_y $ [see Eq.~(\ref{2.1})] are
within the boundary along the $x$-axis, i. e. $-L_x/2 < x_j \le
L_x/2$. It follows that the degeneracy of each Landau level
is $N = L_xL_y B/ \phi_0$, with $\phi_0 = hc/e$.

Consider now the addition of a periodic potential, with a lattice
defined by two non-collinear vectors $\bvec a_1$ and $ \bvec a_2$,
such that $V(\bvec r) = V(\bvec r + n \bvec a_1 + m\bvec a_2)$ for any
 $n, m \in {\cal Z}$. The periodic potential has non-vanishing Fourier
components only at the reciprocal lattice vectors $\bv g = h \bv g_1 +
k \bv g_2$, where $\bv g_i\cdot\bv a_j = 2 \pi \delta_{ij}$ and $h,k$
are integers. Thus:
\begin{equation}
\EqLabel{2.4} V(\bvec r) = \sum_{\bvec g} V_{\bvec g} e^{i\bvec{
r\cdot g}}.
\end{equation}
Further, since $V(\bvec r)$ is real, it follows that $V_{\bvec g} =
V^*_{-\bvec g}$. 

In the absence of Landau level mixing, the Hofstadter spectrum for
both square\cite{Hofst}
\begin{equation}
\EqLabel{2.6} V_s(x,y) = 2A\left[ \cos{2 \pi \over a}x +\cos{2 \pi
\over a}y \right],
\end{equation}
and triangular\cite{Wannier}
$$ V_t(x,y) = -2A\left[\cos {4 \pi \over \sqrt{3}a}x + \cos {2 \pi
\over \sqrt{3}a}\left(x -y \sqrt{3} \right) \right.
$$
\begin{equation}
\EqLabel{2.7} \left. + \cos {2 \pi \over \sqrt{3}a}\left(x +y
\sqrt{3} \right)\right]
\end{equation}
periodic potentials, with nonzero Fourier components only for {\it the
shortest reciprocal lattice vectors}, have been studied extensively in
the literature.\cite{Hofst,Wannier,Geisel,Gerhardts} The parameter
defining the spectrum is the ratio between the flux $\phi={\bv
B}\cdot(\bv a_1 \times \bv a_2)$ of the magnetic field through a unit
cell and the elementary flux $\phi_0$. For $\phi/\phi_0=q/p$, where
$p$ and $q$ are mutually prime integers, the original Landau level is
split into $q$ sub-bands.

We would like to emphasize a qualitative difference between the two
types of potentials: the square potential in Eq.~(\ref{2.6}) is
particle-hole symmetric, 
since $V_s(x,y) = -V_s(x+{a\over 2},y+{a \over 2})$. As a result, the
sign of its amplitude is irrelevant. On the other hand, the triangular
potential does not have this symmetry. With the sign chosen in
Eq. (\ref{2.7}) and $A>0$, $V_t$ has deep local minima at the sites of
the triangular lattice, whereas the maxima are relatively flat and
located on a (displaced) honeycomb lattice. As a result, the sign of
$V_t$ is highly relevant. The second fact that must be mentioned is
that the choice made in Eqs. (\ref{2.6}) and (\ref{2.7}) is rather
simple, since it aligns the periodic potential with the edges of the
sample in a very specific way. In general, however, one could
consider the case where the periodic lattice is rotated by some finite
angle with respect to the sample edges; study of such cases will be discussed
in future work. Finally, it may seem that this choice of periodic
potentials is very restrictive also because only the shortest lattice
vectors have been kept in the Fourier expansion. In fact, the methods
we employ can be directly used for potentials with more Fourier
components, but their inclusion leads to no new physics.

\section{Disorder Potential}
\label{sec3}

Real samples always have disorder. The current consensus is that
high-quality GaAs/AlGaAs samples exhibit a slowly varying, smooth
disorder potential. In a semi-classical picture, the allowed electron
trajectories in the presence of such disorder follow its
equipotential lines.\cite{Springer, 
Trugman} Closed trajectories imply localized electron states, while
extended trajectories connecting opposite edges of the sample are
essential for current transport through the sample (for more details, see
Sec.~\ref{classical}). 

In typical experimental setups,\cite{Sorin} dopant Si impurities with
a concentration of $\sim10^{13}$~cm$^{-2}$ are introduced in a thin
layer of 6~nm in thickness, located 20~nm above the GaAs/AlGaAs
interface. Typically, up to 10\% of the Si atoms are ionized. A small
fraction of the ionized electrons migrate to the GaAs/AlGaAs interface
where they form the 2D electron gas. The electrostatic potential
created by the ionized impurities left behind is the major source of
disorder in the 2DES layer. On the length-scale we are interested in,
there are $10^4$ to $10^5$ such ionized Si impurities per $\mu
m^2$. The resulting disorder potential must be viewed as a collective
effect of the density fluctuation of the ionized
impurities\cite{Nixon} rather than a simple summation of the Coulomb
potential of a few impurities. The electrostatic potential from Si
impurities is compensated and partially screened by other mobile
negative charges in the system such as, for example, the surface
screening effect by mirror charges considered by Nixon and
Davies.\cite{Nixon} An exact treatment of this problem is difficult,
since one should consider the spatial correlation of the ionized
impurities. \cite{Stopa, DasSarma} One model used to describe such
disorder consists of randomly placed Gaussian scatterers.\cite{Ando}
This model captures the main feature of a smooth disorder potential
and supports classical trajectories on equipotential contours, but it has
no natural energy/length scales associated with it. As a result, here
we choose to also investigate a different model of the disorder, which
incorporates the smooth character of the Coulomb potential in real space.

We generate a realization of the disorder potential in the following
way: positive and negative charges, corresponding to a total
concentration of $10^{3}\,\mu m^{-3}$ are randomly distributed within
a volume
$[-L_x/2,L_x/2]\times[-L_y/2,L_y/2]\times[20\mbox{nm}+d,26\mbox{nm}+d]$
above the electron gas which is located in the $z=0$ plane. Here, we
choose $d=4$~nm as an extra spacer since the electronic wave-functions
are centered about 3-5~nm below the GaAs/AlGaAs interface. Since we
are not simulating single impurities but density fluctuations, these
charges are not required to be elementary charges. Instead, we use a
uniform distribution in the range $[-e,e]$ for convenience (a Gaussian
distribution would also be a valid choice), and sum up all Coulomb
potentials from these charges, using the static dielectric constant in
GaAs $\epsilon = 12.91$.\cite{Ralph} The resulting disorder potential
has energy and length scales characteristic of the real
samples. Typical contours for such potentials are shown in
Sec.~\ref{sec5}.

In an infinite system, in the quantum Hall regime, the existence of 
quantum Hall steps implies the existence of critical energies at which
the localization length diverges.\cite{Halperin} This is the quantum
analog of the two dimensional percolation problem in a smooth random
landscape, for which there exists a single critical
energy.\cite{Trugman} In the case of potentials with electron-hole symmetry
$\left<V({\bvec r})\right>=0$, the critical energy lies in
the middle of the band ($E_c=0$), leading to percolating path at half
filling. For a finite mesoscopic sample, however, not only does the
percolating path (critical energy) $E_c$ deviate from this value, but
in samples without a periodic boundary condition one need not have a
percolating path traversing the system in the desired direction. This
arises from the fluctuations near the edge of a mesoscopic system with
free boundary conditions.

We circumvent such a possibility by adding an extra smooth potential
$V'(x,y)$ to the impurity-induced disorder potential $V_i(x,y)$, such
that the total potential $V = V_i + V'$ is zero on the opposite edges
$ x= \pm L_x/2$ of the sample where the metallic leads are
attached. The supplementary contribution $V'(x,y)$ can be thought of
as simulating the effect of the leads on the disorder potential, since
the metallic leads hold the potential on each edge constant by
accumulating extra charges near the interface. Therefore, physically
we expect that the extra potential $V'$ decays exponentially over the
screening length $\lambda$ inside the sample. This implies:
$$ V'(x,y) = -{V_i\left(-{L_x/2},y\right)+V_i(L_x/2, y) \over 2 }
{\cosh(x/\lambda)\over \cosh(L_x /2\lambda)}
$$
$$ +{V_i(-L_x/2,y)-V_i(L_x/2,y) \over 2 }
{\sinh(x/\lambda)\over\sinh(L_x /2\lambda)}
$$ where $\lambda$ is taken to be $100$~nm in our calculation.

\begin{figure}[htbp]
  \centering \includegraphics[width=\FigWidth]{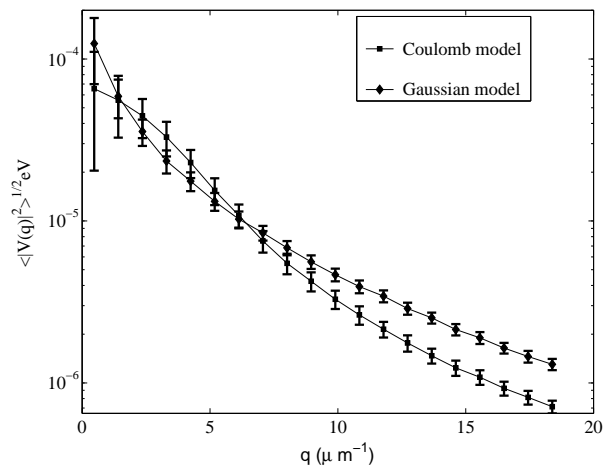}
  \caption{Averaged Fourier amplitudes of two types of disorder
  potential as a function of wavevector $q = |{\bvec q}|$. For both
  Coulomb and Gaussian model, $V(q)^2$ is averaged over 116 disorder
  realizations. The relation between $V(q)$ and $V(\bvec r)$ and
  relevant parameters are discussed in the text. The standard deviation,
  $(L_xL_y)^{-1} \left<\int d\bvec{r} V^2(\bvec r)\right>$ for the Coulomb
  model is $3.2\times10^{-7}$eV$^2$, and $2.1\times10^{-7}$eV$^2$ for the
  Gaussian model.}
  \label{fig1.5}
\end{figure}

In Fig.~\ref{fig1.5}, we plot the average of Fourier transform of the
magnitude of the random potential $\sqrt{\left< |V(q)|^2\right>}$
versus $q = |{\bvec q}|$ for the Coulomb model and the Gaussian model. The Gaussian
model is generated by adding 100 randomly placed Gaussian scatterers
on an area of $3\mu m \times 3\mu m$, each contributing $A_d
e^{-r^2/d^2}$, where $A_d$ is uniformly distributed in
$[-2,2]$~meV, and $d$ is uniformly distributed in
$[0,0.2]$~$\mu$m. $V(\bvec q)$ is related to $V(\bvec r)$ by
$V(\bvec{r}) = \sum_{\bvec q} V(\bvec q) e^{i{\bvec q \cdot r}}$,
where the summation is over all the wavevectors involved in the fast
Fourier transformation. The Gaussian model has an arbitrary energy
scale which is fixed by the maximum value of the distribution
$A_m$. Here $A_m = 2$~meV. As can be seen, $V(q)$ of both models are
decreasing functions of $q$. The trends of decay are exponential at
large $q$. At small $q$, the two models behave differently. Despite the
difference, both models lead to the same qualitative results, although,
as expected, minor quantitative differences are present. This shows
that the physics we uncover is independent of the particular type of
slowly-varying disorder potential considered, and therefore should be
relevant for the real samples.

\section{Numerical Calculations}
\label{sec4}

In this section we discuss the numerical methods we use, including
derivations of some relevant formulas. As already stated, we focus on
the case where the amplitudes of the periodic and disorder potentials
are very small compared to the cyclotron energy and the Zeeman
splitting, and therefore inter-level mixing is ignored.

\subsection{Semi-classical treatment}
\label{classical}

The semi-classical approach is valid\cite{Trugman} for the integer
quantum Hall effect in the presence of a slowly varying, smooth
disorder potential and large magnetic fields (such as we consider), so
that the magnetic length $l$ which determines the spatial extent of
the electron wave-functions is much smaller than the length scale of
variation of the smooth disorder potential, $|\nabla V(\bv r)| \ll
\hbar\omega_c/l$. Then, semi-classically the electron moves along the
equipotential contours of the disorder potential $V(\bv r)$, in the
direction parallel to $\nabla V(\bv r) \times \bvec B $. Since the
kinetic energy is quenched in the lowest Landau level, the total
energy of the electron simply equals the value of the disorder
potential on the equipotential line on which its trajectory is
located. As a result, the density of states in the semi-classical
approach is directly given by the probability distribution for the
disorder potential, which can be calculated by randomly sampling the
potential energy and plotting a histogram of the obtained
values.\cite{Trugman, note1}

In Sec.~\ref{sec5} we compare the results obtained within this
semi-classical approach with fully quantum mechanical results. As
expected, the agreement is good if only the disorder potential is
present. However, if the periodic modulation is also included, the
lattice constant $a$ provides a new length-scale which is comparable
to the magnetic length $l$, and the semi-classical picture breaks
down. Quantum mechanical calculations are absolutely necessary to
quantitatively treat this case.

\subsection{Quantum Mechanical Treatment}

As shown in Sec.~\ref{sec2}, for a finite sample of size $L_x
\times L_y$ at a given magnetic field $B$, the degeneracy of the
unperturbed Landau level is $N = L_xL_yB/\phi_0 = L_xL_y/(2\pi
l^2)$. Since the disorder varies very slowly, we need to consider
systems with $L_x,L_y \gg l$ to properly account for its effects. As a
result, the number of states in a Landau level can be as large as
$10^4$ in our calculations. Storage of the Hamiltonian as a dense
matrix requires considerable amount of computer memory and its direct
diagonalization is prohibitively time-consuming. Sparse matrix
diagonalization techniques could be employed, but they are less
efficient when all eigenvectors are needed, and also
have some stability issues.

Here we introduce the numerical methods we use to compute densities of
states and infer the nature (localized or extended) as well as the
spatial distribution of the wave-functions, while avoiding direct
diagonalization.

\subsubsection{Matrix elements}

Since inter-level mixing is ignored, the Hilbert subspaces
corresponding to different spin-polarized Landau levels do not
hybridize. Each Hilbert subspace $(n,\sigma)$ has a basis described
by Eq.~(\ref{2.1}), containing $N$ orthonormal vectors indexed by
different $k_y$ values.

In order to compute matrix elements of the total Hamiltonian in such a
basis, we use the following identity derived in
Ref. \onlinecite{Gerhardts} (notice their different sign convention
for $k_y$. If $\sigma\ne \sigma'$, the overlap is zero):
\begin{equation}
\EqLabel{2.5} \bra{n', k'_y} e^{i\bvec{q\cdot r}}\ket{n,k_y} =
\delta_{k'_y,k_y-q_y} {\cal L}_{n',n}(\bvec q) e^{{il^2\over
2}q_x(k'_y+k_y)},
\end{equation}
where
\begin{eqnarray*}
{\cal L}_{n',n} (\bvec q) = \left(m! \over M! \right)^{1 \over
        2}i^{|n'-n|}\left[ q_x+iq_y \over
        \sqrt{q_x^2+q_y^2}\right]^{n-n'} && \\ \times e^{-{1 \over 2}
        Q}Q^{{1 \over 2}|n'-n|}L_m^{(|n'-n|)}(Q), &&
\end{eqnarray*}
with $Q = {1 \over 2}l^2(q_x^2+q_y^2)$, $m$ and $M$ the minimum and
the maximum of $n'$ and $n$ respectively, and $L^{(|n'-n|)}_m(Q)$ the
associated Laguerre polynomial. When band-mixing is neglected $n = n'$
and ${\cal L}_{n,n} (\bvec q) = e^{-{1 \over 2}Q} L_n(Q)$. For the
first Landau level, $L_0(x) =1$.

Eq.~(\ref{2.5}) gives us the matrix elements for the square
 [Eq.~(\ref{2.6})] or triangular [Eq.~(\ref{2.7})] periodic
 potentials. In either case, there are Fourier components
 corresponding to $q_y=\pm 2\pi/a$ and $q_y=0$.  Since only basis
 vectors for which the difference $k_y-k_y'=q_y$ give non-vanishing
 matrix elements, we must choose the length $L_y$ of the sample to be
 a multiple integer of $a$, the lattice constant.

The matrix elements of the disorder potential are computed in a
similar way. We use a grid of dimension $N_x\times N_y$ to cover the
sample and generate the values of the disorder potential on this
grid. Then, fast Fourier transform (FFT)\cite{FFT} is used to find the
long wavelength components of the disorder potential corresponding to
the allowed values $q_{x,y}=0,\pm{2\pi\over L_{x,y}}, ..., \pm
\left[{N_{x,y}\over 2}\right]{ 2\pi\over L_{x,y}}$ (proper care is
taken to define Fourier components so that $V_{\bv q} = V^*_{-\bv
q}$). The matrix elements of this discretized disorder potential are
then computed using Eq.~(\ref{2.5}). In principle, finer grids
(increased values for $N_x$ and $N_y$) will improve accuracy. However,
they also result in longer computation times, since they
add extra matrix elements in the sparse matrix, corresponding to large
wave-vectors. We have verified that a grid size of dimension
$N_x=N_y=72$ is already large enough to accurately capture the
landscape of a $3\mu m \times 3\mu m$ sample and the computed
quantities have already converged, with larger grids leading to hardly
noticeable changes. This procedure is also justified on a physical
basis. First, the neglected large wave-vector components describe very
short-range spatial features, which are probably not very accurately
captured by our disorder models to begin with, and which are certainly
not believed to influence the basic physics. Secondly, this procedure
insures that the actual disorder potential we use is periodic in the
$y$-direction, since each Fourier component retained has this
property. This is consistent with our use of a basis of wave-functions
which are periodic along $y$.

The matrix elements of the Hamiltonian within a given Landau level
$(n,\sigma)$ are then $\langle n, k_y,\sigma | {\cal H} | n,
k'_y,\sigma\rangle= E_{n,\sigma} + \langle n, k_y | {\cal V} | n,
k'_y\rangle$, where $E_{n,\sigma}$ are given by Eq.~(\ref{2.2}) and
the matrix elements of both the periodic and the disorder part of the
potential ${\cal V}$ are computed as already discussed. This produces
a sparse matrix, which is stored efficiently in a column compressed
format.

\subsubsection{Densities of States and Filling Factors}
\label{subs2}

A quantity that can be computed without direct diagonalization is the
filling factor $\nu_{n,\sigma}(E_F)$ as a function of Fermi
energy. The filling factor is defined as:
\begin{equation}
\EqLabel{4.6} \nu_{n,\sigma}(E_F) = { 1 \over N} \sum_{\alpha}
\Theta(E_F- E_{n,\alpha,\sigma}),
\end{equation}
where $\Theta(x)$ is the Heaviside function and $N$ is the total
number of states in the $(n,\sigma)$ Landau level. (Since we neglect
Landau-level mixing, we can define this quantity for individual
levels.) The filling factor tells us what fraction of the states in
the given Landau level are occupied at $T=0$, for a given value of the
Fermi energy. This corresponds to the average filling factor measured
in experiment and is also proportional to the integrated total (as
opposed to local) density of states.

The filling factor is straightforward to compute if the eigenenergies
$E_{n,\alpha,\sigma}$ are known. However, we want to avoid the
time-consuming task of numerical brute force diagonalization. The
strategy we follow is a generalization to Hermitian matrices of the
method used in Ref. \onlinecite{czhou}. We restate the problem in the
following way: assume we have a Hermitian matrix of size $N \times N$
(no Landau level mixing), given by the matrix elements of $M = {\cal
H} - E_F {\bf 1} $ in the basis $|n, k_y, \sigma\rangle$ (${\bf 1}$ is
the unit matrix). Then, $\nu_{n,\sigma}(E_F)$ is proportional to the
number of negative eigenvalues of the matrix $M$. We now generate the
quadratic form ${\cal M} = \sum_{i,j = 1}^{N} \zeta_i \zeta_j^*
M_{ij}$, and transform it into its standard form ${\cal M} = \sum_{i=
  1}^{N} d_i|\chi_i|^2$ using the Jacobian method described
below. Here, 
$d_i$'s are all real numbers, and the $\chi_i$'s are linear
combinations of the $\zeta_i$'s. This is a similarity transformation
which retains the signature of the matrix. As a result, even though
the numbers $d_i$ are not eigenvalues of $M$, the number of negative
eigenvalues equals the number of negative $d_i$ values. It follows
that $\nu_{n,\sigma}(E_F)$ is obtained by simply counting the number
of negative $d_i$ values for the given $E_F$.

The Jacobian method is iterative in nature. First, all terms
containing $\zeta_1$ and $\zeta_1^*$ are collected and the needed
complementary terms are added to form the first total square
$d_1|\chi_1|^2$, so that $\zeta_1$ and $\zeta_1^*$ are eliminated from the rest of the quadratic form ${\cal M}$. The procedure is then repeated for all $\zeta_2$ and
$\zeta_2^*$ terms (producing $d_2$) etc., until all $N$ values $d_i$
are found. Computationally, this can be done by scanning the lower or
upper triangle of the Hermitian matrix $M$ only once. The total number
of operations is proportional to the number of nonzero elements of the
matrix, meaning that for a dense matrix it scales with $N^2$ (sparse
matrices require much fewer operations). As a result, this procedure
is much faster than brute force diagonalization which scales with
$N^3$ (for us, $N\sim 10^4$). The filling factor $\nu_{n,\sigma}(E)$
is a sum of step-like functions, with steps located at the
eigenvalues. By scanning $E$ and identifying the position of these
steps we can also find the true eigenvalues $E_{n,\alpha,\sigma}$,
with the desired accuracy. Finally, the total density of states is
given by $ \rho_{n\sigma}(E) = d \nu_{n,\sigma}(E)/dE$.

\subsubsection{Green's functions: extended vs. localized states}
\label{subs3}

The advanced/retarded Green's functions are the solutions of the
operator equation
\begin{equation}
\EqLabel{4.1} \left(\hbar\omega - {\cal H} \pm i \delta\right) {\hat
G}^{R,A}(\omega) = {\bf 1},
\end{equation}
where $\delta \rightarrow 0^{+}$. (In practice we use a set of small
positive numbers, and use the dependence on $\delta$ to obtain
results.) If the exact eigenstates and eigenvalues of the total
Hamiltonian ${\cal H}$ are known, ${\cal H} | n, \alpha, \sigma\rangle
= E_{n,\alpha,\sigma} | n, \alpha, \sigma\rangle, $ (no Landau level
mixing), it follows:
\begin{equation}
\EqLabel{4.2} {\hat G}^{R,A}(\omega) = \sum_{n,\alpha,\sigma}^{}{ |
n,\alpha,\sigma\rangle\langle n, \alpha, \sigma| \over \hbar\omega -
E_{n,\alpha,\sigma} \pm i\delta} = \sum_{n,\sigma}^{} {\hat
G}^{R,A}_{n,\sigma}(\omega).
\end{equation}
The exact eigenstates can be expanded in terms of the basis states
$|n, k_y,\sigma\rangle$ as
\begin{equation}
\EqLabel{4.3} | n, \alpha, \sigma\rangle = \sum_{k_y}
c_{n,\alpha}(k_y) |n, k_y,\sigma\rangle.
\end{equation}
Since the states $|n, k_y,\sigma\rangle$ are localized near $x=k_yl^2$
[see Eq.~(\ref{2.1})], the coefficients $c_{n,\alpha}(k_y)$ describe
the probability amplitude for an electron in the state $| n,
\alpha, \sigma\rangle$. 
Knowledge of these coefficients allows us to infer whether
such states are extended or localized in the $x$-direction,
i.e. whether they can carry currents between the leads.

However, as already stated, we wish to avoid direct
diagonalization. We can still infer whether the Hamiltonian has
extended or localized wave-functions near a given energy $\hbar\omega$
in the following way. We introduce the matrix elements:
$$ G^{R,A}_{n,\sigma}(k_y, k'_y; \omega) = \langle n, k_y,\sigma |
{\hat G}^{R,A}(\omega) | n, k'_y,\sigma\rangle
$$
\begin{equation}
\EqLabel{4.4} = \sum_{\alpha}^{} { c_{n,\alpha}(k_y)
c^*_{n,\alpha}(k'_y) \over \hbar\omega - E_{n,\alpha,\sigma} \pm
i\delta}.
\end{equation}
If Landau level mixing is neglected, Eq.~(\ref{4.1}) can be rewritten
in the basis $|n, k_y,\sigma\rangle$ as:
$$ \sum_{k''_y}^{} \left[(\hbar\omega \pm i\delta) \delta_{k_y,k''_y}
  - \langle n, k_y,\sigma | {\cal H} | n, k''_y,\sigma\rangle \right]
$$
\begin{equation}
\EqLabel{4.5} \times G^{R,A}_{n,\sigma} (k''_y,k'_y;
\omega)=\delta_{k_y,k'_y}.
\end{equation}

We use the popular numerical library SuperLU,\cite{SuperLU} based on
LU decomposition and Gaussian reduction algorithm for sparse matrices,
to solve these linear equations. Consider now the matrix element $
G^{R,A}_{n,\sigma}(k_{\min},k_{\max}; \omega) $ corresponding to the
smallest $k_y=k_{\min}$ and the largest $k_y=k_{\max}$ values. If all
wave-functions with energies close to $\hbar\omega$ are localized in
the $x$-direction, it follows that
$|G^{R,A}_{n,\sigma}(k_{\min},k_{\max}; \omega)| $ is a very small
number, of the order $e^{-L_x/\xi(\omega)}$, where $\xi(\omega)$ is
the localization length at the given energy. On the other hand, we
expect to see a sharp peak in the value of
$|G^{R,A}_{n,\sigma}(k_{\min},k_{\max}; \omega)| $ if $\hbar\omega$ is
in the vicinity of an extended state eigenvalue, since [see
Eqs. (\ref{4.3},\ref{4.4})] both $c_{n,\alpha}(k_{\min})$ and
$c_{n,\alpha}(k_{\max})$ are non-vanishing for an extended
wave-function with significant weight near both the $-L_x/2$ and the
$L_x/2$ edges. Moreover, the height of this peak scales like
$1/\delta$, so by varying $\delta$ we can easily locate the energies
of the extended states.

\subsubsection{Green's functions: local densities of states}
\label{subs4}

We can also use Green's functions techniques to image the local
density of states at a given energy $E$. By definition (and neglecting
Landau level mixing), the local density of states in the level
$(n,\sigma)$ is:
$$ \rho_{n,\sigma}(\bv r;E) = \sum_{\alpha}^{} |\langle \bv r| n,
\alpha, \sigma\rangle |^2 \delta\left(E-E_{n,\alpha,\sigma}\right)
$$
\begin{equation}
\EqLabel{5.1} = { 1 \over \pi} \mbox{Im}\langle \bv r| {\hat
 G}^{A}_{n,\sigma}(E)|\bv r\rangle,
\end{equation}
where the second equality follows from Eq.~(\ref{4.2}). This function
traces the contours of probability $|\phi_{n,\alpha,\sigma}(\bv r)|^2$
for electrons with the given energy $E$. Its direct computation,
however, is difficult and very time-consuming.

For the rest of this subsection, the discussion is restricted to the
Lowest Landau level $n=0$ (the value of $\sigma$ is irrelevant). We
know that in the lowest Landau level, electronic wave-functions cannot
be localized in any direction over a length-scale shorter that the
magnetic length $l$. As a result, it suffices to compute a projected
local density of states on a grid with $l\times l$ (or larger)
spacings. The projection is made on maximally localized
wave-function, defined as follows. Let $\bv r_0=(x_0,y_0)$ be a point
on the grid. We associate it with a vector:
\begin{equation}
\EqLabel{5.2} |x_0, y_0\rangle = \sum_{k_y}^{} |k_y\rangle \langle
k_y| x_0,y_0\rangle,
\end{equation}
where we use the simplified notation $|k_y\rangle \equiv
|n=0,k_y,\sigma\rangle$ for the basis states of the first Landau level
(see Eq.~(\ref{2.1})) and we take
\begin{equation}
\EqLabel{5.3} \langle k_y| x_0,y_0\rangle = \sqrt{{2l\pi^{1 \over2}}
\over L_y} e^{-{x_0^2 \over 2l^2} - {k_y^2l^2\over 2} +
k_y(x_0+iy_0)}.
\end{equation}
It is then straightforward to show that
\begin{equation}
\EqLabel{5.4} \langle \bv{r}| x_0, y_0\rangle = { 1\over \sqrt{2\pi}l}
e^{- {(x-x_0)^2 \over 4l^2 } - {(y-y_0)^2 \over 4l^2 }}e^{-{i \over
2l^2}(x+x_0)(y-y_0)}.
\end{equation}
In other words, $|x_0,y_0\rangle$ is an eigenstate of the first Landau
level strongly peaked at $\bv r= \bv r_0$. (The phase factor is due to
the proper magnetic translation). We then define the projected density
of states [compare with Eq.~(\ref{5.1})]:
\begin{equation}
\EqLabel{5.5} \rho_P (x_0,y_0; E) = { 1 \over \pi} \mbox{Im} \langle
x_0,y_0|\hat{G}^{A}(E)|x_0,y_0\rangle,
\end{equation}
and use it to study the spatial distribution of the electron wave-functions
at different energies. Strictly speaking, the local density
of states defined in Eq.~(\ref{5.1}) cannot be projected exactly on
the lowest Landau level, because the lowest Landau level does not
support a $\delta$-function ($\langle \bv{r}| n,k_y,\sigma\rangle \ne
0$, $\forall n$). However, the coherent states $|x_0,y_0\rangle$ we
select are the maximally spatially-localized wave functions in the
lowest Landau level, and have the added advantage that they can be
easily stored as sparse vectors, because of their Gaussian profiles
[see Eq.~(\ref{5.3})]. Moreover, in the limit $l \rightarrow 0$ ($B
\rightarrow \infty$) where $|\langle\bv r | x_0,
y_0\rangle|\rightarrow \delta(x-x_0)\delta(y-y_0)$, the projected
density of states $\rho_P(x_0,y_0; E)\rightarrow
\rho_{0,\sigma}(\bv{r}; E)$. Therefore, for the large $B$ values that
we consider here, the projected density of states $\rho_P$ should
provide a faithful copy of the local density of states.

We compute the projected local density of states following the method
of Ref. \onlinecite{Haydock}. Let $\bv u_0$ be the vector with
elements $\langle k_y| x_0, y_0\rangle$ obtained from the
representation of $|x_0,y_0\rangle$ in the $|k_y\rangle$ basis [see
Eq.~(\ref{5.2})], and let $H$ be the matrix of the Hamiltonian ${\cal
H}$ in the $|k_y\rangle$ basis. We generate the series of orthonormal
vectors $\bv u_0, \bv u_1, ...$ using:
\begin{eqnarray*}
        \bvec{v}_1 &=& H \bvec{u}_0 ,\\ a_0 &=& \bvec{u}_0^\dagger
        \bvec{v}_1 ,\\ \bvec{u}_1 &=& { \bvec{v}_1 - a_0\bvec{u}_1
        \over \sqrt{\bvec{v}_1^\dagger \bvec{v}_1-a_0^2}},
\end{eqnarray*}
and for $n \ge 2$,
\begin{eqnarray*}
        \bvec{v_n} &=& H \bvec{u}_{n-1} ,\\ a_{n-1} &=&
        \bvec{u}_{n-1}^\dagger \bvec{v_n} ,\\ b_{n-2} &=&
        \bvec{u}_{n-2}^\dagger \bvec{v_n} ,\\ \bvec{u}_n &=&
        {\bvec{v}_n - a_{n-1} \bvec{u}_{n-1} - b_{n-2}\bvec{u}_{n-2}
        \over \sqrt{\bvec{v}_n^\dagger \bvec{v}_n - a_{n-1}^2 -
        b_{n-2}^2 }}.
\end{eqnarray*}
The numbers $a_n$ and $b_n$ can be shown to be real. We do not have a
``terminator''\cite{Haydock} to end this recursive series. Instead, our procedure
ends when the orthonormal set of vectors $\bv u_0, \bv u_1, ...$
exhausts a subspace of the lowest Landau level containing all states
coupled through the disorder and/or periodic potential to the state
$|x_0,y_0\rangle$ (i.e., all states that contribute to the projected
DOS at this point). In the presence of disorder, this usually
includes the entire lowest Landau level.

Then, the projected density of states is given by Eq.~(\ref{5.5}),
where the matrix element of the Green's function is the continued
fraction:
$$ \langle x_0,y_0|G^A(E)|x_0,y_0\rangle=$$
\begin{equation}
\EqLabel{5.6} \left[E-i\delta-a_0 -b_0^2\left[E-i\delta - a_1
        -b_1^2\left[ \ldots\right]^{-1} \right]^{-1} \right]^{-1}
\end{equation}
Because the Hamiltonian is a sparse matrix, the generation of these
orthonormal sets and computation of $\rho_p(E)$ for all the grid
points is a relatively fast procedure. Moreover, this computation is
ideally suited for parallelization, with different grid points
assigned to different CPUs.

\section{Numerical Results}
\label{sec5}

In this section we present numerical results obtained using these
methods. We have analyzed over 20 different disorder realizations for
samples of different sizes, and all exhibit the same qualitative
physics. Here, we show results for several typical samples. The
lattice constant is always $a=39$~nm if periodic potential is present,
as defined by the experimental system.\cite{Sorin}

\begin{figure}[t]
\includegraphics[width=\FigWidth]{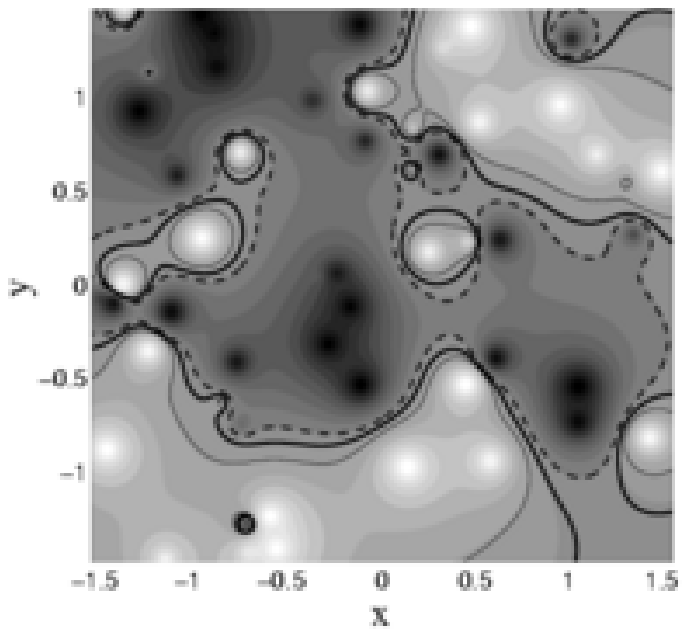}
\includegraphics[width=\FigWidth]{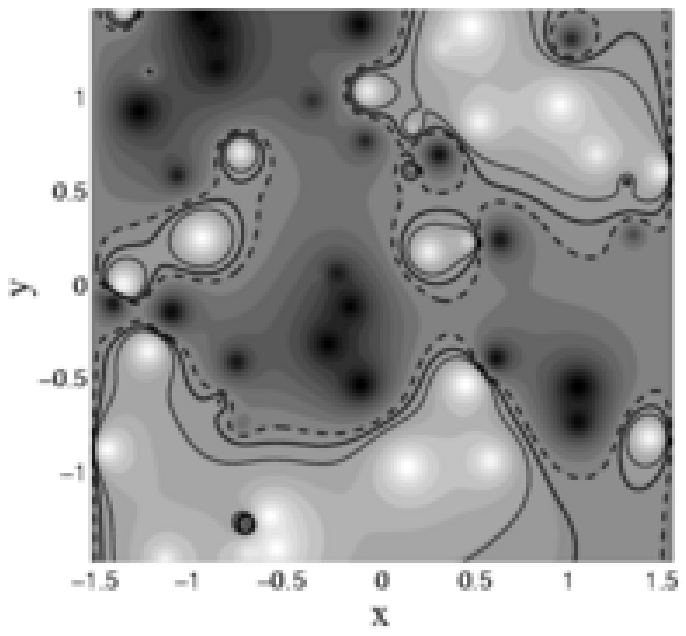}
\caption{\label{fig2} Profile of the disorder potential obtained from
our Coulomb model on a $3.11\mu$m$\times 2.96\mu$m sample, without
(upper panel) and with (lower panel) the $V'(\bv r)$ correction at the
$x=\pm L_x/2$ edges. The disorder potential varies between $-3$~meV
and 3~meV, on a spatial length-scale much larger than
$l=12.03$~nm. The critical region containing extended states is in the
vicinity of $E=0.06$~meV. The contours are shown for $E=$0.0575~meV
(dashed), 0.17~meV (thick solid) and 0.31~meV (thin solid). These
energy values correspond to classical filling factors $\nu$=0.47, 0.58
and 0.68 in the upper panel and $\nu$=0.45, 0.56, 0.66 in the lower
panel. The difference is due to the supplementary smooth potential $V'$.}
\end{figure}

\begin{figure}[t]
\includegraphics[angle=270,width=\FigWidth]{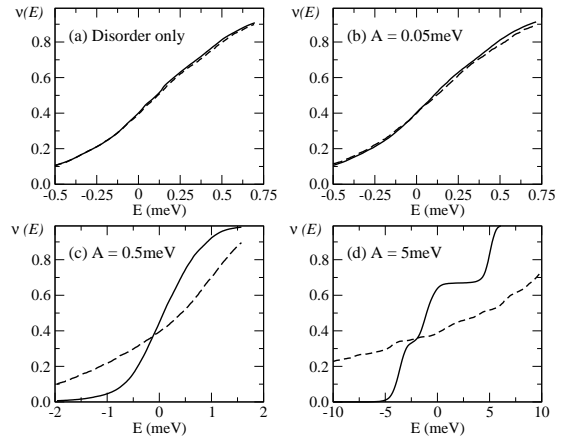}
\caption{ Semi-classical (dashed line) and quantum (solid line) filling
  factors for the disorder potential shown in
  Fig.~\ref{fig2}, but different amplitudes of the triangular periodic
  potential (a) $A$=0; (b) $A$=0.05meV; (c) $A$=0.5meV and (d) $A$=5
  meV. As expected, agreement exists only in the limit $A \rightarrow
  0$. }
\label{fig3}
\end{figure} 

\begin{figure}[hbp]
\includegraphics[angle=270,width=\FigWidth]{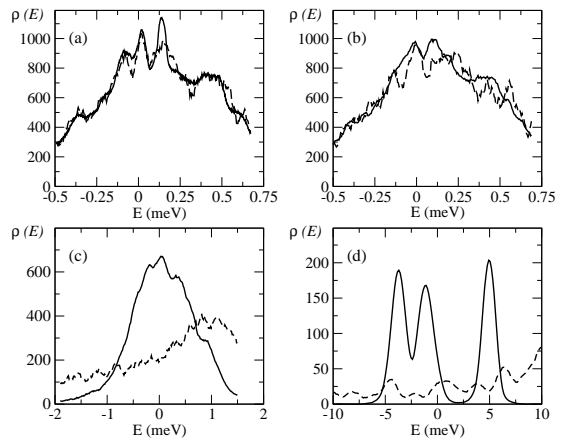}
\caption{ Semi-classical (dashed line) and quantum (full line) density
  of states calculated from corresponding filling factors in Fig.
  \ref{fig3}. We show only the center of the disorder-broadened lowest
  Landau level, where the density of states is large. }
\label{fig4}
\end{figure} 

For the first sample, we consider $\phi/\phi_0=3/2$ ($B =
4.71$~T). The magnetic length is $l= 12.03$~nm, and we choose a sample 
size $L_x=3.11\mu$m and $L_y= 76a=2.964\mu$m. With these choices, the
Landau level contains $N=10108$ states. The disorder potential
obtained with our scheme described in Sec.~\ref{sec3} is shown in
Fig.~\ref{fig2}, both with and without the correction $V'(\bv r)$. An
extended equipotential line appears, as expected, at 
$\nu \approx 0.5$.

In Figs. \ref{fig3} and \ref{fig4} we plot the filling factor $\nu(E)$
and the corresponding total density of states $\rho(E)$ as a function
of $E$ (computation details were given in Sec.~\ref{subs2}). These
quantities are obtained in the semi-classical limit (dashed line) and
with the full, quantum-mechanical treatment (solid line). Results are
shown for 4 different cases: (a) only disorder potential and (b, c, d)
disorder plus a triangular periodic potential with amplitudes
$A=0.05$, 0.5 and 5~meV, respectively. We only plot a relatively small
energy interval where the DOS is significant, and ignore the
asymptotic regions with long tails of localized states.

While the agreement between the semi-classical and quantum-mechanical
treatment is excellent in the limit $A\rightarrow 0$, the two methods
give more and more different results as the periodic potential
amplitude is increased. As already explained, this is a consequence of
the fact that the magnetic length $l$ is comparable to the lattice
constant $a$, leading to a failure of the semi-classical treatment
when this extra length-scale is introduced. In particular, in the case
with the largest periodic potential [panel (d) of Figs.~\ref{fig3} and
\ref{fig4}] we can clearly see the appearance of the 3 subbands
expected for the Hofstadter butterfly at $\phi/\phi_0=3/2$, although
the disorder leads to broadened and smooth peaks, and partially
fills-in the gap between the lower two subbands. This picture [panel
(d)] is quite similar to the density of states that
Ref.~\onlinecite{MacDonald} calculated using the self-consistent Born
approximation. This is expected since the SCBA approach is valid in
the limit of strong periodic potential with weak disorder. However,
the SCBA approach is not appropriate in the limit of moderate or
strong disorder, where the higher order terms neglected in SCBA are no
longer small. For disorder varying on a much longer length-scale
than the periodic potential, one still expects that {\it locally}, on
relative flat regions of disorder, the system exhibits the
Hofstadter-type spectrum. However, these spectra are shifted with
respect to one another by the different local disorder values. If
disorder variations are small, then the total spectrum shows somewhat
shifted subbands with partially filled-in gaps, but overall the
Hofstadter structure is still recognizable. On the other hand, for
moderate and large disorder, the detailed structure of the local
density of states from various flat regions are hidden in the total
density of states. All one sees are some broadened, weak peaks and
gaps superimposed on a broad, continuous density of states.

\begin{figure}[t]
\includegraphics[width=\FigWidth]{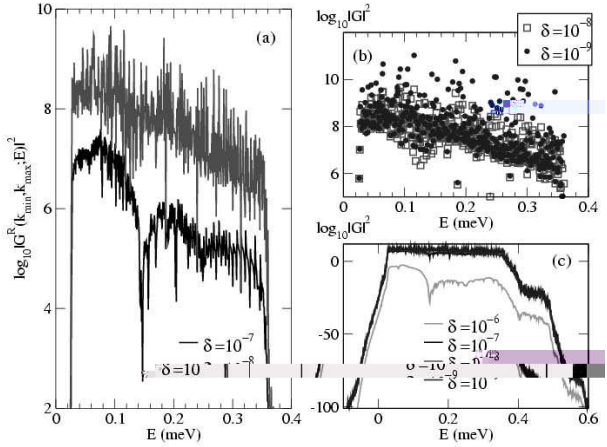}
\caption{ Semi-log plot of the amplitude of Green's function matrix
  element between the two edge states near $x=\pm L_x/2$, as a
  function of energy. Only the disorder potential of Fig.~\ref{fig2}
  is present. (a) comparison between $\delta=10^{-7}$ and
  $\delta=10^{-8}$ results; (b) comparison between $\delta=10^{-8}$
  and $\delta=10^{-9}$ results; (c) comparison between results
  corresponding to $\delta =\delta=10^{-6},10^{-7},10^{-8}$ and
  $10^{-9}$. (the last three curves are indistinguishable to the eye
  on this scale.) All $\delta$ values are in eV units. }
\label{fig5}
\end{figure}

We now analyze the nature of the electronic states for these
configurations. We start with the case which has only disorder. In
Fig.~\ref{fig5} we plot $|G^R(k_{\min},k_{\max};E)|^2$ as a function
of the energy $E$, for different values of $\delta$ (computation
details were given in Sec.~\ref{subs3}). As already discussed,
extended states are indicated by large values of this quantity, as
well as a strong (roughly $1/\delta^2$) dependence on the value of the
small parameter $\delta$.

Figure \ref{fig5} reveals that as $\delta$ is reduced, resonant
behavior appears in a narrow energy interval $E= 0.02-0.36$~meV. Panel
(a) shows that results corresponding to $\delta = 10^{-7}$~eV and
$\delta = 10^{-8}$~eV indeed differ by roughly 2 orders of magnitude,
with $\delta = 10^{-8}$~eV showing sharper resonance peaks. The
difference between results for $\delta = 10^{-8}$~eV and $\delta
=10^{-9}$~eV shown in panel (b), is no longer so definite. The reason
is simply that for such small $\delta$, the denominator in the Green's
function expression is usually limited by $|E-E_{n,\alpha,\sigma}|$
and not by $\delta$ [see Eq.~(\ref{4.4})], and the dependence on
$\delta$ is minimal. Only if $E$ is such that
$|E-E_{n,\alpha,\sigma}| <\delta$ can we expect to see a $\delta$
dependence, and indeed this is observed at some energies. Finally, in
panel (c) we show the comparison with a larger energy interval. The
value of the Green's function decreases exponentially fast on both sides
of the critical region, indicating strongly localized states. Here,
data for $\delta = 10^{-6}$~eV is a smooth curve, whose magnitude is
much less than that of the other three values even for localized
states. This is due to the fact that this $\delta$ is larger than
typical level spacings. As a result, several levels contribute
significantly to Green's function at each $E$ value,
and the destructive interference of
the random phases of different eigenfunctions lead to the
supplementary $\delta$-dependence. We conclude that the disorder
potential has a critical energy regime of approximately $0.3$~meV width,
covering less than 5\% (in energy) and 20\% (in number of states) of
the disorder-broadened band with total width $\sim 6$ meV. The
position of the critical energy interval is in agreement with the
semi-classical results which suggest an extended state in the vicinity
of $E=0.06$meV. By comparison with Fig.~\ref{fig3}, we can also see
that this critical regime corresponds to a roughly half-filled band,
in agreement with the experiment.

\begin{figure}[t]
\includegraphics[width=\FigWidth]{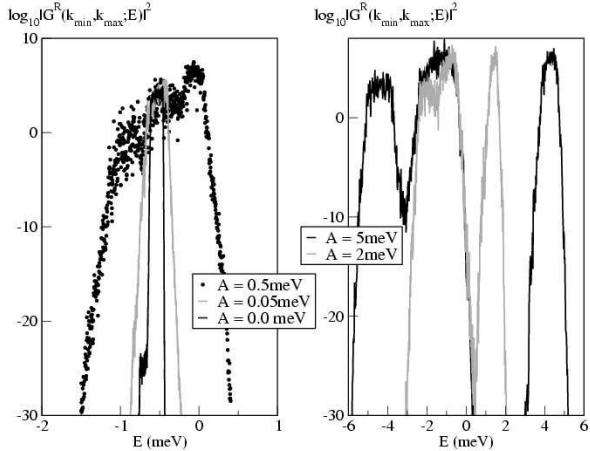}
\caption{ The effect of a triangular periodic potential on the
critical energy regime. The disorder potential used here (not shown)
supports a narrow interval of extended states centered at about
$-0.6$~meV. As the amplitude $A$ of the periodic potential increases,
the range of extended states increases dramatically. The left panel
shows results for disorder-only and two relatively weak periodic
potentials, while the right panel shows two larger periodic potentials
where the three-subband structure expected for $\phi/\phi_0 = 3/2$ is
clearly seen.}
\label{fig6}
\end{figure}

The effect of an additional triangular periodic potential is shown in
Fig.~\ref{fig6}, where we plot the same quantity shown in
Fig.~\ref{fig5} for a fixed $\delta = 10^{-7}$~eV and different
amplitudes $A=0$, 0.05, 0.5 and 5~meV, respectively. These results
correspond to a different Coulomb disorder potential (not shown), as
can be seen from the different location of its extended states. Here
we see how the narrow critical interval of extended states grows
gradually as the amplitude of periodic potential is increased and
finally exhibits the three well-separated extended subbands expected
for $\phi/\phi_0 = 3/2$ in the limit of vanishing disorder. The three
subbands can already be resolved for the moderate case $A = 0.5~$meV,
although they are very wide and exhibit significant overlap.

\begin{figure}[t]
\includegraphics[width=\FigWidth]{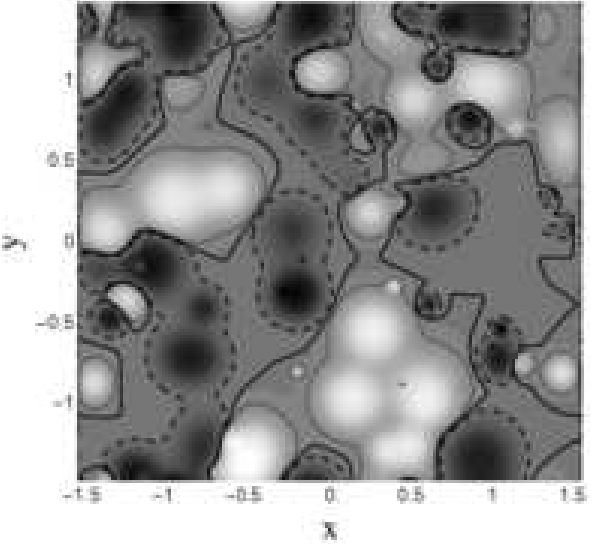}
\caption{A disorder potential of Gaussian type on a roughly $3\mu m
\times 3\mu m$ square. The three lines are equipotential contours
close to the critical regime, with energies of -0.1~meV (dashed),
0~meV (thick solid) and 0.1~meV (thin solid). Cyclic boundary
condition are applied in the $y$ direction.}
\label{fig7bis}
\end{figure}

\begin{figure}[t]
\includegraphics[width=\FigWidth]{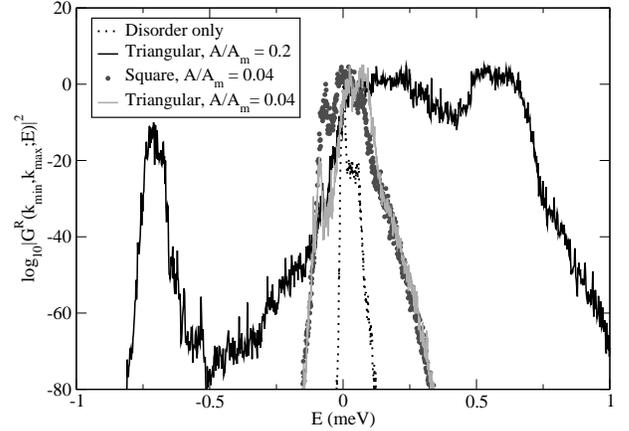}
\caption{ Green's functions for a sample with Gaussian disorder and
various periodic potentials. The calculation included 20216 states
with $\phi/\phi_0=3$. Similar to results shown in Fig.~\ref{fig7}, we
see that the periodic potentials widen the critical region.}
\label{fig7}
\end{figure}

Qualitatively similar behavior is obtained if we use the Gaussian
scatterers model for disorder. A typical realization of this disorder is shown in
Fig.~\ref{fig7bis}. Results for the Green's function's values with such
disorder are shown in Fig.~\ref{fig7}, for cases with pure disorder,
and also cases with either a triangular or a square periodic
potential. The magnetic field has been doubled, such that
$\phi/\phi_0=3$. Similar to the case shown in Fig.~\ref{fig6}, the
periodic potential leads to a widening of the critical regime. For
large periodic potentials, the expected Hofstadter-like three-subband
structure emerges again.  We conclude that Coulomb and Gaussian
disorder models show qualitatively similar behavior.

We now analyze the projected local density of states $\rho_P(E)$
discussed in Sec.~\ref{subs4}, in order to understand the reason for
this substantial widening of the critical region by even small
periodic potentials. We consider a smaller sample, of size
approximately $1.6\mu$m$\times1.6\mu$m, and compute the projected
density of states for 500 equally-spaced energy values, on a
$60\times60$ square grid and for a value $\delta = 10^{-8}$~eV. This
$\delta$ value is comparable or smaller than the level spacing, so we
expect to see sharp resonances from the contribution of individual
eigenfunctions as we scan the energy spectrum. Each computation
generates a large amount of data (roughly 24M), corresponding to the
500 plots of the local density of states at the 500 values of
$E$. Since we  cannot show all this data, we select a couple of
representative cases and some statistical data to interpret the
overall results.

\begin{figure}[t]
\includegraphics[width=\FigWidth]{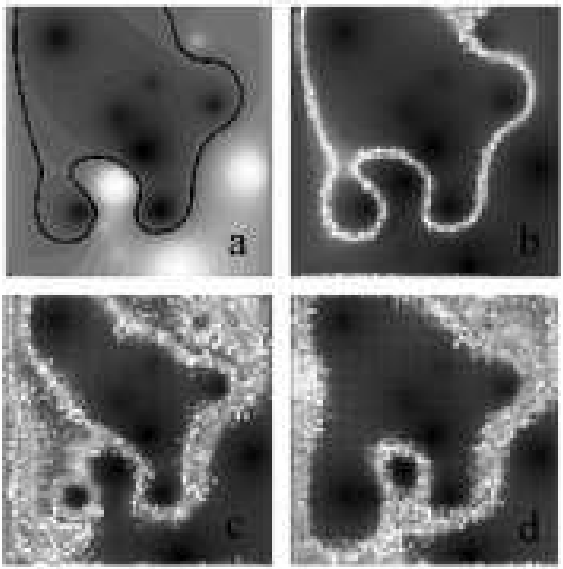}
\caption{ (color online) Projected local density of states 
  $\rho_P(\bv r;E)$ for
  $E=-0.504$~meV. Panel (a) shows the profile of the disorder
  potential, and the equipotential contour (black line) corresponding
  to $E=-0.504$~eV. The other three panels show $\rho_P(\bv r;E)$
   for (b)  
  disorder only; (c) disorder plus triangular periodic potential with
  $A = 0.1$~meV; (d) disorder plus square periodic potential with $A =
  0.1$~meV. The width and length of the sample are both $1.6\,\mu$m, 
  and $\phi/\phi_0=3/2$. Increased brightness corresponds to  larger
  values.}
\label{fig9}
\end{figure}

\begin{figure}[t]
\includegraphics[width=\FigWidth]{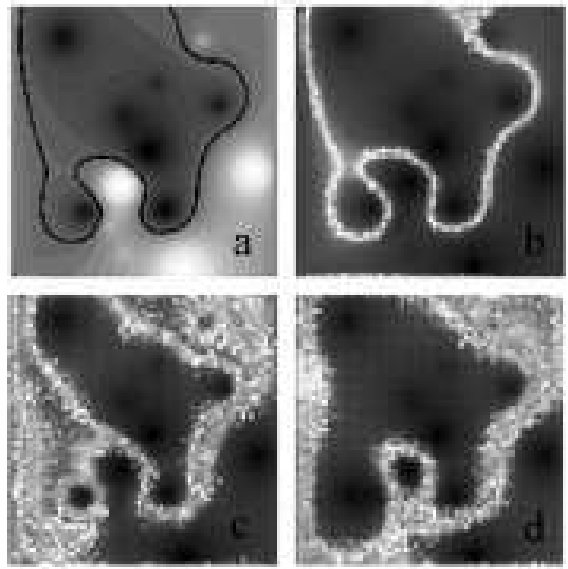}
\caption{ (color online) The same as in Fig.~\ref{fig9}, but for an energy
$E=-0.124$~meV close to the band center. }
\label{fig10}
\end{figure}

Figures \ref{fig9} and \ref{fig10} show some of our typical
results. The two figures are calculated for the same Coulomb-disorder
potential, for values of $E=-0.504$~meV (at the bottom of the band)
and $E=-0.124$~meV (close to, but below the band center) respectively.
Each figure contains four panels, panel (a) shows the profile of the
disorder potential as well as an equipotential line (solid black)
corresponding to the value $E$ considered; the other  three panels show
the projected density of states $\rho_P(E)$ for (b) pure disorder; (c)
disorder plus triangular periodic potential with $A=0.1$~meV; (d)
disorder plus square periodic potential with $A=0.1$~meV. In
Fig.~\ref{fig9}, this equipotential line (which traces the
semi-classical trajectory of electrons with the same energy $E$)
surrounds local minima of the disorder potential, suggesting localized
electron states at such low energies. Indeed, this is what panels (b),
(c) and (d) show. The projected density of states $\rho_P(E)$ is large
(bright color) at the positions where electrons of energy $E$ are
found with large probabilities. For pure disorder, we observe only
closed trajectories (localized states), whose shape is in excellent
agreement with the semi-classical trajectory, as expected. If a
moderate periodic potential is added, the wave-functions spread over a
larger area, and nearby contours sometimes merge together. Instead of
sharp lines, as seen in panel (b), the contours now show clear
evidence of interference effects of the wave-functions on the periodic
potential decorating the electron reservoirs. Some periodic modulations can
also be observed in the background of panels (c) and (d), especially
for the square potential. These are {\em not } the direct oscillations
of the periodic potentials, since the grid we use to compute these
figures has a linear size equal to $7/10$ of the period $a=39$~nm of
the periodic potential. Capturing detailed behavior inside each unit
cell would require a much smaller grid, which is not only time
consuming, but also violates the requirement that the grid size be of
order $l$ or larger.

Figure \ref{fig10} for an energy close to the band center, shows the
same characteristics. For pure disorder, the electrons at this energy
 trace a sharp contour very similar 
to the corresponding equipotential line shown in panel (a). Electrons are
still not delocalized, since this contour does not connect either pair
of opposite edges. However, addition of the periodic potential now
leads to extended states for both types of periodic potentials [(c)
and (d)] at roughly $E=-0.124$~meV, 
demonstrating the widening of the critical region with the
addition of a periodic potential.

Physically, one can understand this spread of the wave function in the
presence of the periodic potential using the semi-classical
picture.\cite{Sorin} If only a smooth disorder potential is present,
the equipotential at any energy $E$ must be a smooth, continuous
line. However, if a periodic potential with minima $-V_m$ and maxima
$V_M$ is superimposed over disorder, the new equipotential line now
breaks into a series of small ``bubbles'' surrounding the
disorder-only contour. This happens throughout the area defined by the
equipotentials $E-V_M$ and $E+V_m$ of the disorder potential, since
the addition of the periodic potential leads new regions in this area
to have a total energy $E$. Quantum mechanically, we expect some
tunneling inside this wider area and this is indeed what we observe in
Figs.~\ref{fig9} and \ref{fig10}. This mechanism suggests {\em
enhanced delocalization } on both sides of the critical region as
localized wave functions spread out over larger areas, as well as a
widening of the critical region itself, in agreement with our
numerical results.

This spreading of the wave functions in the presence of the periodic
potential can also be characterized by counting, at a given energy
$E$, the number of grid points $\bv r$ which have a value $\rho_P(\bv
r; E) > \rho_c$, where $\rho_c$ is some threshold value. For sufficiently large
$\rho_c$, this procedure counts grid points where electrons with
energy $E$ are found with large probabilities, thus, in effect it
characterizes the ``spatial extent'' of the wave functions. 
\begin{figure}[t]
\includegraphics[width=0.8\FigWidth,angle=-90]{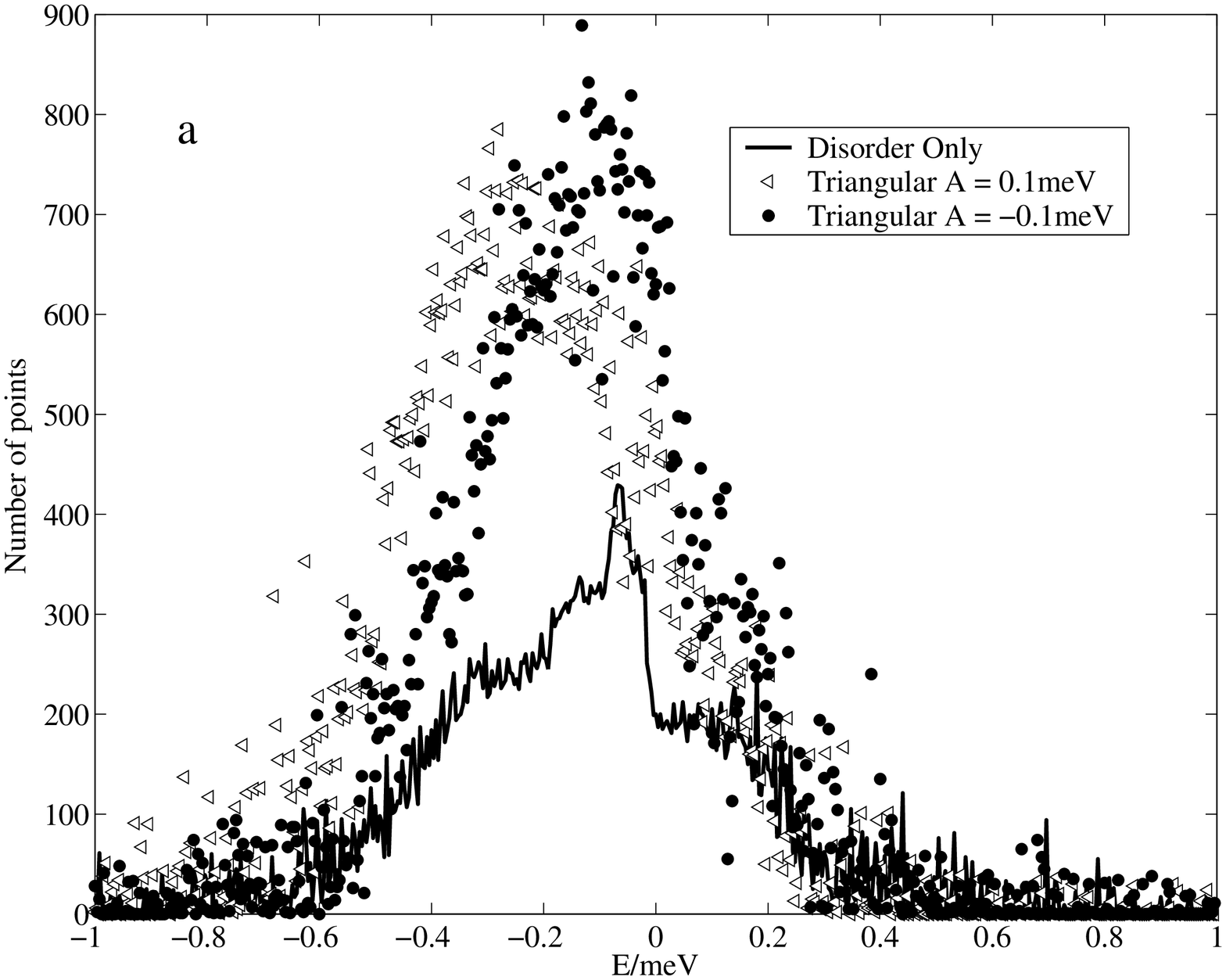}
\includegraphics[width=0.8\FigWidth,angle=-90]{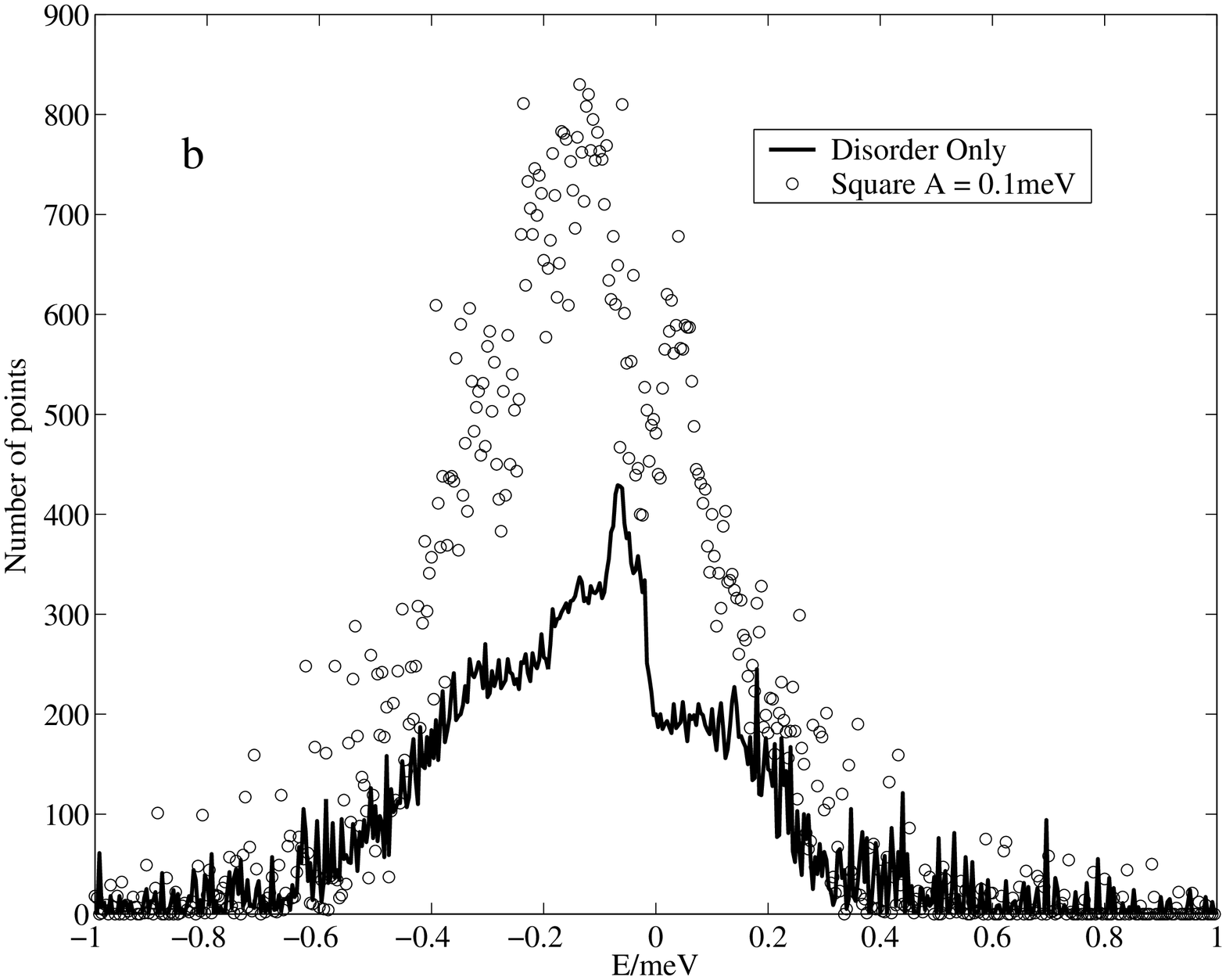}
\caption{ Number of grid points for which $\rho_P(\bv r, E)>100$ as a
  function of the energy $E$. This quantity characterizes the
  ``spatial extent'' of the wave function. The upper panel compares
  results for disorder only (A=0) and disorder plus triangular
  potentials with $A \pm 0.01$~meV. The difference observed for the
  two signs is a consequence of particle-hole asymmetry of the
  triangular potential. The lower panel shows results for disorder
  only and disorder plus a square potential.  }
\label{fig11}
\end{figure}
The results of such counting are shown in Fig.~\ref{fig11} for 500
energy values corresponding to the disorder potential analyzed in
Figs. \ref{fig9} and \ref{fig10}. There are a total of
$60\times60=3600$ grid points on the sample. For the case of pure
disorder (black line) we see that the largest values are found at
energies just below 0, where the extended states (the critical region)
are found for this particular realization of disorder. Because it is a
smooth, sharp line, even the most extended trajectory has significant
probabilities at only about 10\% of the grid sites. For both higher
and lower energies, this number decreases very fast, indicating wave
functions localized more and more about maxima or minima of the
disorder potential, as expected. Addition of a small periodic
potential increases this number substantially, clearly showing the
supplementary spreading of the wave functions in the presence of the
periodic potential.

Figure \ref{fig11} shows this effect for three types of periodic
potential: triangular lattices with $A > 0$ and $A<0$ (upper panel),
and square lattice in the lower panel. All three cases show
significant enhancement, as compared to the pure
disorder case. In addition, we see that while the square potential
gives a fairly symmetrical enhancement, the triangular potential does
not, with curves for $\pm A$ not overlapping. This is a consequence of
the asymmetric shape of the periodic potential, which has different
values for its minima and maxima $|V_m|\ne |V_M|$, as well as
different arrangements for the points where minima/maxima appear
(triangular lattice vs. honeycomb lattice). Fig.~\ref{fig11} clearly
shows that $A >0$ favors increased delocalization below the critical
energy regime, while $A <0$ favors increased delocalization above it.

\begin{figure}[t]
\includegraphics[width=0.9\FigWidth,angle=-90]{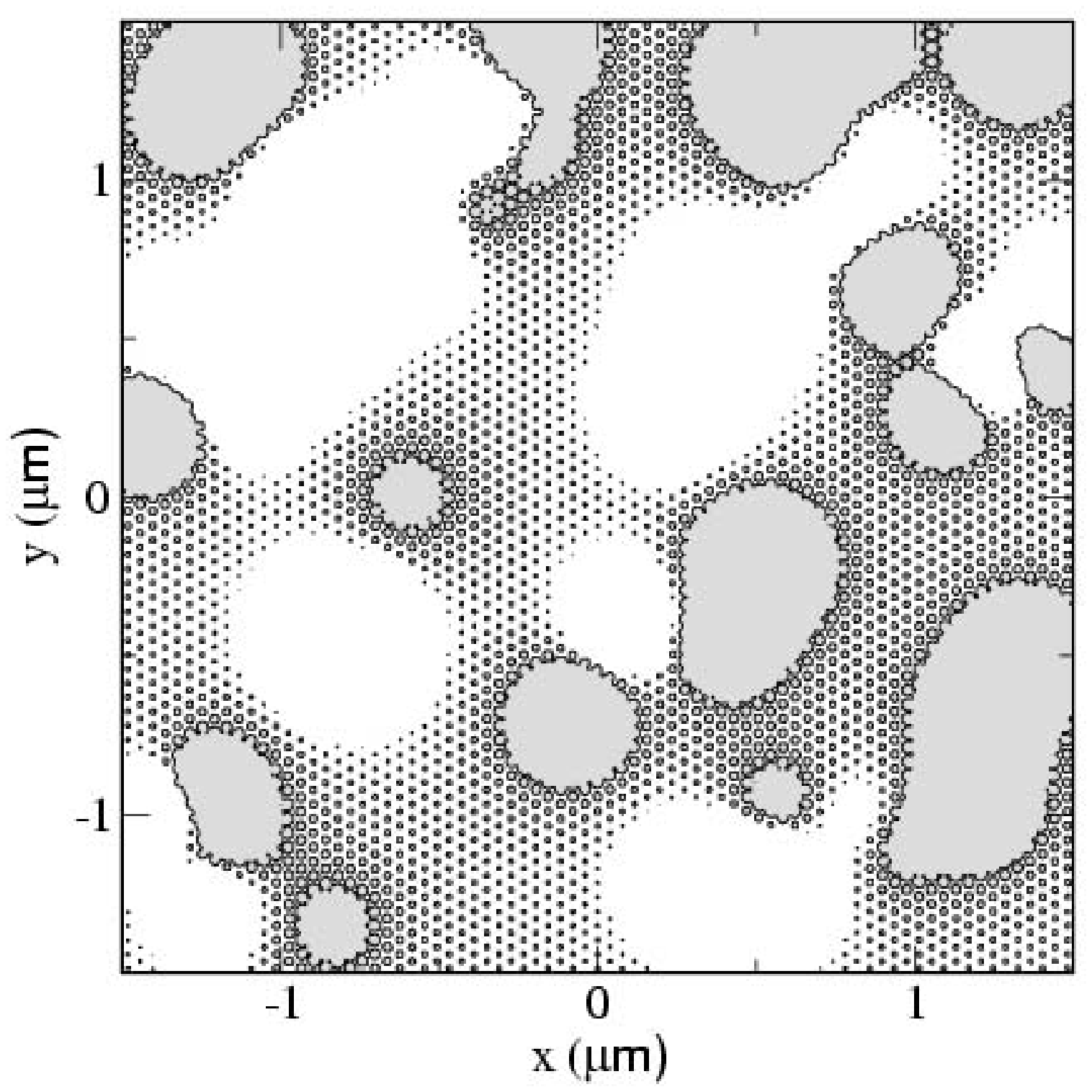}
\includegraphics[width=0.9\FigWidth,angle=-90]{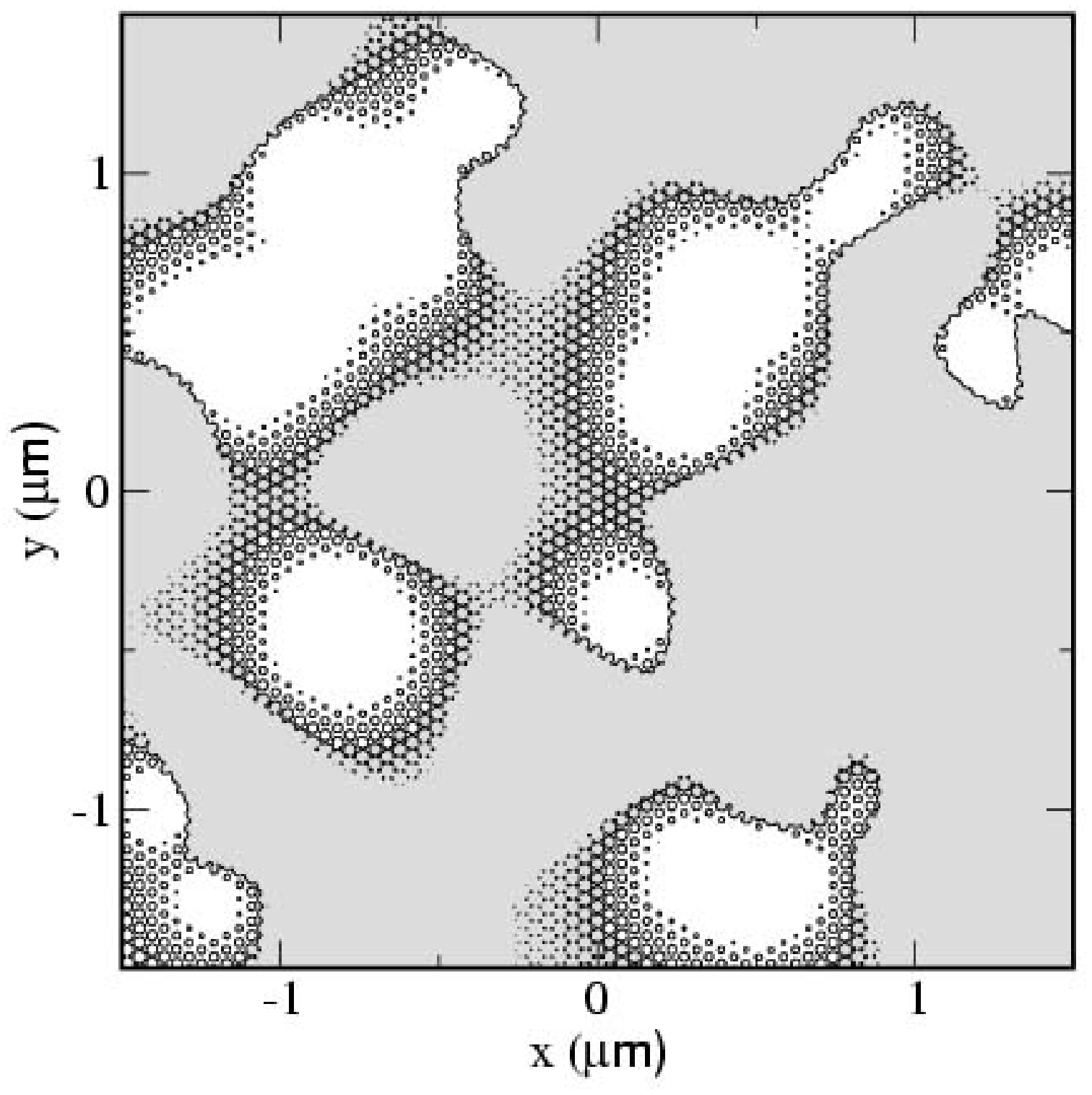}
\caption{ Equipotential contours at filling factor $\nu = 0.3$ (upper
panel) and $\nu = 0.7$ (lower panel) for a $3 \mu m \times 3 \mu m$
sample with disorder plus a small triangular periodic potential with
$A >0$. The shaded regions correspond to energies below the
respective equipotential. In the semi-classical approximation, the
shaded areas are filled with electrons, with the maximum density of
$1/2\pi l^2$, whereas the white areas are completely devoid of
electrons. Quantum-mechanically, one expects interference in the
regions with small periodic ``bubbles'', induced by the periodic
potential (see Figs. \ref{fig9} and \ref{fig10}).}
\label{fig12}
\end{figure}

The reason for this different response to the two signs of the triangular
potential can be nicely explained within the semi-classical
framework. In Fig.~\ref{fig12} we show the equipotential lines corresponding to
filling factors $\nu=0.3$ (well below critical region) and $\nu=0.7$
(well above the critical region) for a realization of Coulomb disorder
(not shown) plus a triangular potential with $A>0$. Areas with energy
 below the equipotential value are shaded. In this
case we can clearly see that instead of the continuous, smooth
trajectory expected for disorder-only cases, there are also extra
``bubbles'' regions connecting the areas between such contours. Since
the choice $A>0$ leads to deep minima at $-V_m=-6A$ with triangular
arrangement and relatively flat maxima at $+V_M=3A$ with honeycomb
arrangement [see Eq.~(\ref{2.7})], it follows that the triangular
(honeycomb) ``bubbles'' region appear roughly in the area bounded by
the equipotentials $E$ and $E+V_m$ (respectively, $E-V_M$ and $E$) of
the pure disorder potential. At low filling factors, the pure disorder
$E$ equipotential is a collection of closed contours surrounding local
minima [see panel (d) of Fig.~\ref{fig9} for an illustration]. It
follows that for the choice $V_m > V_M$, the more extended region with
triangular ``bubbles'' will be found outside these ``islands'' and
will lead to a spread of the wave function over considerably larger
areas, as indeed seen in the upper panel of Fig.~\ref{fig12}. On the
other hand, at large filling factors the disorder-only contours are
``islands'' surrounding the maxima of the disorder potential. In this
case, contours between $E$ and $E+V_m$ are inside the $E$ contour, so
the triangular ``bubbles'' region does not help to connect various
``islands'' as before. The honeycomb ``bubbles'' regions does this,
but because $V_M < V_m$ the extension of the wave function between
``islands'' is significantly smaller in this case.

In the quantum-mechanical case one expects interference (due to
tunneling) effects between the small ``bubbles'' regions, and
therefore a wave function which is extended over their entire area, as
indeed we observe to be the case in Figs.~\ref{fig9} and
\ref{fig10}. In other words, one expects that a triangular potential
with $A>0$ will lead to considerable increase of the localization
length, and respectively widening of the critical energy region, at
filling factors below one-half, whereas $A<0$ will favor
delocalization at filling factors above one-half, as seen in
Fig.~\ref{fig11}. This asymmetry is therefore clearly a consequence
of the asymmetry of the triangular potential, and is absent for a
square potential with only lowest order Fourier coefficients, which
possesses electron-hole symmetry. This should have clear implications
for the transport properties of the system.

\section{Summary and Conclusions}
\label{sec6}

In this study we have investigated the effects of moderate-to-large
smooth disorder on the Hofstadter butterfly expected for 2DES in a
perpendicular magnetic field and a pure periodic modulation. The
parameters of our study are chosen so as to be suitable for the
interpretation of recent experiments on a two dimensional electron
system in GaAs/AlGaAs heterostructure with periodic modulation
provided by a diblock copolymer, \cite{Sorin} The experiment shows
that (i) the longitudinal resistance $R_{xx}$ is still peaked
approximately at half filling; (ii) there are many reproducible oscillations in
$R_{xx}$, indicating non-trivial electronic structures in the
patterned sample; (iii) the distribution of these oscillatory features
is asymmetric, with most of them appearing on the high magnetic fields
(i.e. low filling factors $\nu < 0.5$) side of the peak of $R_{xx}$;
and (iv) the temperature dependence of $R_{xx}$ indicates that the
asymmetric off-peak resistance is thermally excited, whereas the
central $R_{xx}$ peak (close to half filling) has metallic behavior.

These observations cannot be explained on the basis of the Hofstadter
structure.\cite{Sorin} This is not surprising, since one expects that
large disorder will modify this structure considerably. Effects of
small disorder on the Hofstadter butterfly had been investigated
previously using SCBA,\cite{MacDonald} but this basically
perturbational approach is not appropriate for the case of
moderate-to-large smooth disorder. Instead, we identify and use a number of
techniques which give the exact solution (if electron-electron
interactions, as well as inelastic scattering are neglected) while
avoiding brute force numerical diagonalizations.

Our results demonstrate that while the Hofstadter butterfly is
destroyed by large disorder, the effects of the periodic potential are
non-trivial for states near the critical regime. Firstly, they lead to
a significant increase in localization lengths of the localized states
at mesoscopic ($\mu m$) length scale and induce an effective
widening of the critical regime near the critical
regime. This is achieved through a spreading of the electron wave-function 
on the flat regions of the slowly varying disorder potential,
where their behavior is dominated by the periodic modulation. This
regime shows an interesting transition between the pure disorder and
the pure periodic potential cases. In the case of pure disorder, the
semi-classical approach tells us that at finite filling factors, some
areas of the sample are fully occupied by electrons with the maximum
possible density of $1/(2\pi l^2)$ (these are the areas where the
disorder potential has minima) whereas other areas are fully devoid of
electrons (areas where the potential has maxima) and the boundary
between such regions is very sharp. On the other hand, for a pure
periodic modulation all wave functions have translational invariance
with the proper symmetry, and therefore electron densities are uniform
over the entire sample (up to small periodic modulations inside each
unit cell).

When both types of potential are present, with disorder being
dominant, our results show {\it three} types of areas. There are
regions which are fully occupied and regions which are completely
devoid of electrons, as in the case of pure disorder. However, the
periodic potential leads to a widening of the boundary between the
two, where the wave functions interact with several oscillations of
the periodic modulation and therefore have some partial local
filling. As the critical regime containing wave functions percolating
throughout the sample is approached, this spreading of the wave
function becomes dominant in establishing the transport properties of
the system.

An equivalent way to say this is that the main effect of the periodic
potential is to provide bridging between the fully-occupied electron
``puddles'' created by the disorder potential. Since the connecting
areas are relatively flat, the local wave functions respond to the
local periodic potential, and therefore locally have a
Hofstadter-butterfly like structure. If the partial filling factor in
such a region is inside the gap of the local Hofstadter butterfly
structure, one expects no transport through this local area. This
should result in a dip in the longitudinal transport, since in such
cases the periodic potential will not transport electrons from one
``puddle'' to another one. By contrast, if the local filling factor in
such a region is inside a subband of a local Hofstadter structure,
this area will establish a link between different ``puddles'' and thus
help enhance the transport through the sample. Transport in this
regime should show strong thermal activated behavior, in contrast to
metallic transport in the critical regime where the wave functions
connect opposite edges of the sample.

As a result, one expects a series of local minima and maxima in the longitudinal
resistivity on either side of the central peak induced by the extended
states (critical regime). Furthermore, for an asymmetric triangular
potential, this response should be strongly asymmetric, with the
effect most visible on one side of the central peak. (One must keep in
mind that since tunneling leads to exponential dependencies, even
small differences in the extent of the wave functions can have rather
large effects on $\rho_{xx}$). Such an asymmetry should also be
present in longitudinal conductance of finite but low temperature,
e. g. in the hopping regime which is sensitively dependent on the
nature of the localized wavefunctions, as is indeed seen
experimentally.\cite{Sorin}

To summarize, our qualitative explanation for the various experimental
features are as follows:

(i) The $R_{xx}$ peak is roughly at the center of the band because the
weak periodic potential cannot establish a Hofstadter-like structure
over the whole band. Instead, low and high $\nu$ states are
strongly localized and do not transport longitudinal currents.

(ii) New extended states induced by the periodic potential are
responsible for the
reproducible peaks and valleys appearing in $R_{xx}$.

(iii) The periodic potential also leads to the expansion of localized
wave functions, which contribute to the thermally activated conduction
at lower filling factors. The detailed structure of the wave functions
gives rise to the oscillations of the off-peak $R_{xx}$, similar to
conductance fluctuations.\cite{Shayegan}
Finally, 

(iv) the asymmetry in $R_{xx}$ is a
manifestation of the asymmetry of the triangular potential, which has
a stronger effect at low filling factors than at high filling factors for
$A>0$. We predict that this asymmetry should be absent for a
symmetric square periodic potential.

The weak point in our calculation is that we are unable to accurately
model the potential in the real samples, because various screening
effects have not been properly taken into account. Also, we have no
quantitative information about the magnitude of the periodic potential
in the 2DES layer, because of the additional
strain\cite{Larkin} contribution induced by the periodic decoration. As a
result, we only claim qualitative agreement with the experiment,
although our investigations show the same type of behavior for various
types of disorder potentials and various (small-to-moderate) strengths
of the periodic potential. The most direct check of this work would be
an experimental demonstration that thermally activated conduction
appears symmetrically on both sides of the $R_{xx}$ peak for a
periodic potential with square symmetry and primarily lowest Fourier
coefficients.
 
Limited computer resources restrict our calculations to samples no
larger than $3 \mu$m$ \times 3 \mu $m, while the sample used in
experiment has a size of $20 \mu$m$ \times 20\mu$m. From a
theoretical point of view, it is interesting to ask what is the
thermodynamic limit. For pure disorder, it is believed that in this
limit the typical size of wavefunction diverges at a single critical
energy. Since we cannot pursue size-dependent analysis for samples
larger than $3 \mu$m$ \times 3 \mu $m, we do not know whether the
small periodic potential will lead to a finite size critical regime in the 
thermodynamic limit,
although this seems likely. From an experimental point of view, the
interesting question is whether the Hofstadter structure can be
observed at all. Our studies suggest that this may be possible for
small mesoscopic samples, where the slowly-varying disorder has less
effect. Alternatively, one must find a way to boost the strength of
the periodic modulations inside the 2DES.

\section*{Acknowledgements}

We thank Sorin Melinte, Mansour Shayegan, Paul M. Chaikin and Mingshaw
W. Wu for valuable discussions. We also thank Prof. Li Kai's group in
Computer Science Department of Princeton University for sharing their
computer cluster with us. This research was supported by NSF grant
DMR-213706 (C.Z. and R.N.B.) and NSERC (M.B.). M.B. and R.N.B. also
acknowledge the hospitality of the Aspen Center for Physics, where
parts of this work were carried out.


\begin{thebibliography}{99}

\bibitem{IQHE} K. von Klitzing, G. Dora and M. Pepper,
Phys. Rev. Lett. {\bf 45}, 494 (1980).

\bibitem{FQHE} D. C. Tsui, H. L. St$\ddot o$rmer, and A. C. Gossard,
Phys. Rev. Lett. {\bf 48}, 1559 (1982).

\bibitem{Springer} For a review, see {\it ``The Quantum Hall
Effect''}, edited by R.E. Prange and S.M. Girvin, Graduate Texts in
Contemporary Physics (Springer-Verlag, New York, 1987).

\bibitem{Hofst} D. R. Hofstadter, Phys. Rev. B {\bf 14}, 2239 (1976).

\bibitem{equiv} Dieter Langbein, Phys. Rev. {\bf 180}, 633 (1969); the
 electronic structure in the asymptotic cases is periodic in
 $\phi/\phi_0$ or $\phi_0/\phi$, and the equality is meant modulo this
 periodicity.

\bibitem{germans} D. Springsguth, R. Ketzmerick, and T. Geisel,
Phys. Rev. B {\bf 56}, 2036 (1997).

\bibitem{Holland} T. Schl\"osser, K. Ensslin, J. P. Kottahaus and
M. Holland, Europhys. Lett. {\bf 33}, 683 (1996).

\bibitem{Weimann} D. Weiss, M. L. Roukes, A. Menschig, P. Grambow,
K. von Klitzing and G. Weimann, Phys. Rev. Lett. {\bf 66}, 2790
(1991).

\bibitem{Weiss}C. Albrecht, J. H. Smet, K. von Klitzing, D. Weiss,
V. Umansky and H. Schweizer, Phys. Rev. Lett. {\bf 86}, 147 (2001).

\bibitem{Wulf} Rolf R. Gerhardts, Dieter Weiss and Ulrich Wulf,
Phys. Rev. B {\bf 43}, 5192 (1991).

\bibitem{Sorin} S. Melinte, M. Berciu, C. Zhou, E. Tutuc,
  S. J. Papadakis, C. Harisson, E. P. De Poortere, M. Wu, P. M. Chaikin,
  M. Shayegan, R. N. Bhatt and R. A. Register,  to
  appear in Phys. Rev. Lett. (cond-mat/0311400).

\bibitem{Sorin2} S. Melinte, E. Grivei, V. Bayot, and M. Shayegan,
Phys. Rev. Lett. {\bf 82}, 2764 (1999), and references therein.

\bibitem{Chaikin} P. Chaikin (private communication).

\bibitem{MacDonald} U. Wulf and A. H. MacDonald, Phys. Rev. B {\bf
47}, 6566 (1993).

\bibitem{Huckestein} B. Huckstein and R. N. Bhatt, Surface Science
{\bf 305}, 438 (1994).

\bibitem{Wannier} F.H. Claro and G.H. Wannier, Phys. Rev. B {\bf 19},
6068 (1979).

\bibitem{Geisel} D. Springsguth, R. Ketzmerick, and T. Geisel,
Phys. Rev. B {\bf 56}, 2036, (1997).

\bibitem{Gerhardts} D. Pfannkuche and R. R. Gerhardts, Phys. Rev. B
{\bf 46}, 12606 (1992).

\bibitem{Trugman} S. A. Trugman, Phys. Rev. B {\bf 27}, 7539 (1983).

\bibitem{Nixon} John A. Nixon and John H. Davies, Phys. Rev. B {\bf
41}, 7929 (1990).

\bibitem{Stopa} M. Stopa, Phys. Rev. B {\bf 53}, 9595 (1996); Physica
B {\bf 227}, 61 (1996).

\bibitem{DasSarma} S. Das Sarma and S. Kodiyalam, Semiconductor
Science and Technology {\bf 13}, A59 (1998).

\bibitem{Ando} T. Ando, J. Phys. Soc. Japan {\bf 53}, 3101 (1984).

\bibitem{Ralph} Ralph Williams, {\it ``Modern GaAs Processing
Methods''}, (Artech House Publishers, Boston$\cdot$London, 1990).

\bibitem{Halperin} B. I. Halperin, Phys. Rev. B {\bf 25}, 2185 (1982).

\bibitem{note1} Allowed trajectories are such that the total flux
through the area enclosed by the trajectory is an integer number of
elementary fluxes $\phi_0$. This can be understood in the spirit of
the Bohr-Sommerfeld quantization rule, since it ensures constructive
interference of the wave function around the contour. Except for the
most localized states, found at the bottom and the top of the band,
all other localized states are such that they enclose large numbers of
magnetic fluxes. Imposing the exact quantization condition (which is
numerically time consuming) leads to a negligible change in the value
of the allowed equipotential value with respect to a randomly chosen
value. Since such small changes do not influence the shape of the
density of states, we ignore imposing this quantization condition in
obtaining the semi-classical densities of states.

\bibitem{FFT} we use the package FFTW2.1.3 available on-line at
  http://www.fftw.org.

\bibitem{czhou} Chenggang Zhou and R. N. Bhatt, Phys. Rev. B {\bf 68}, 045101
  (2003).

\bibitem{SuperLU}Xiaoye S. Li, M. Baertschy, T. N. Rescigno,
W. A. Issacs and C. W. McCurdy, Phys. Rev. A {\bf 63}, 022712
(2001). Details regarding the software also available at
http://www.nersc.gov/~xiaoye/SuperLU.

\bibitem{Haydock}Roger Haydock, Phys. Rev. B {\bf 61}, 7953 (2000).

\bibitem{Shayegan} J. A. Simmons, H. P. Wei, L. W. Engel, D. C. Tsui
and M. Shayegan, Phys. Rev. Lett. {\bf 63}, 1731 (1989).

\bibitem{Larkin} J. H. Davies and I. A. Larkin, Phys. Rev. B {\bf 49},
  4800 (1994).

\end{thebibliography}
\end{document}